\documentclass[twocolumn,
superscriptaddress,
nature,
article,
preprintnumbers,
amsmath,amssymb]{revtex4-1}

\usepackage{lipsum}
\usepackage{graphicx,epsf,epsfig,ulem,verbatim}
\usepackage{floatrow,color,soul,lineno}
\usepackage{graphicx,epsf,epsfig,ulem,verbatim}
\usepackage{floatrow,color,soul,lineno}

\usepackage{epsfig}
\usepackage{epstopdf} 
\usepackage{amsmath}
\usepackage[shortlabels,inline]{enumitem}
 \setenumerate{label=\arabic*)}

\usepackage{bm}
\usepackage{subfigure}
\usepackage{color}
\usepackage{natbib}
\setcitestyle{super}

\newcommand*{\citen}[1]{%
  \begingroup
    \romannumeral-`\x 
    \setcitestyle{numbers}%
    \cite{#1}%
  \endgroup   
}

\begin{document}

\title{Valley dependent anisotropic spin splitting in silicon quantum dots}

\author{Rifat Ferdous}
\affiliation{Network for Computational Nanotechnology, Purdue University, West Lafayette, IN 47907, USA}

\author{Erika Kawakami}
\affiliation{QuTech and Kavli Institute of Nanoscience, TU Delft, Lorentzweg 1, 2628 CJ Delft, The Netherlands}

\author{Pasquale Scarlino}
\affiliation{QuTech and Kavli Institute of Nanoscience, TU Delft, Lorentzweg 1, 2628 CJ Delft, The Netherlands}

\author{Micha{\l} P. Nowak}
\affiliation{QuTech and Kavli Institute of Nanoscience, TU Delft, Lorentzweg 1, 2628 CJ Delft, The Netherlands}
\affiliation{AGH University of Science and Technology, Faculty of Physics and Applied Computer Science, al.~Mickiewicza 30, 30-059 Krak\'{o}w, Poland}

\author{D. R.  Ward}
\affiliation{University of Wisconsin-Madison, Madison, Wisconsin 53706, USA}

\author{D. E. Savage}
\affiliation{University of Wisconsin-Madison, Madison, Wisconsin 53706, USA}

\author{M. G. Lagally}
\affiliation{University of Wisconsin-Madison, Madison, Wisconsin 53706, USA}


\author{S. N. Coppersmith}
\affiliation{University of Wisconsin-Madison, Madison, Wisconsin 53706, USA}

\author{Mark Friesen}
\affiliation{University of Wisconsin-Madison, Madison, Wisconsin 53706, USA}

\author{Mark A. Eriksson}
\affiliation{University of Wisconsin-Madison, Madison, Wisconsin 53706, USA}

\author{Lieven M. K. Vandersypen}
\affiliation{QuTech and Kavli Institute of Nanoscience, TU Delft, Lorentzweg 1, 2628 CJ Delft, The Netherlands}

\author{Rajib Rahman}
\affiliation{Network for Computational Nanotechnology, Purdue University, West Lafayette, IN 47907, USA}

\date{\today}

\maketitle

\section*{Abstract}
{Spin qubits hosted in silicon (Si) quantum dots (QD) are attractive due to their exceptionally long coherence times and compatibility with the silicon transistor platform. To achieve electrical control of spins for qubit scalability, recent experiments have utilized gradient magnetic fields from integrated micro-magnets to produce an extrinsic coupling between spin and charge, thereby electrically driving electron spin resonance (ESR). However, spins in silicon QDs experience a complex interplay between spin, charge, and valley degrees of freedom, influenced by the atomic scale details of the confining interface. Here, we report experimental observation of a valley dependent anisotropic spin splitting in a Si QD with an integrated micro-magnet and an external magnetic field. We show by atomistic calculations that the spin-orbit interaction (SOI), which is often ignored in bulk silicon, plays a major role in the measured anisotropy. Moreover, inhomogeneities such as interface steps strongly affect the spin splittings and their valley dependence. This atomic-scale understanding of the intrinsic and extrinsic factors controlling the valley dependent spin properties is a key requirement for successful manipulation of quantum information in Si QDs.}

\begin{figure*}[htbp]
\includegraphics[]{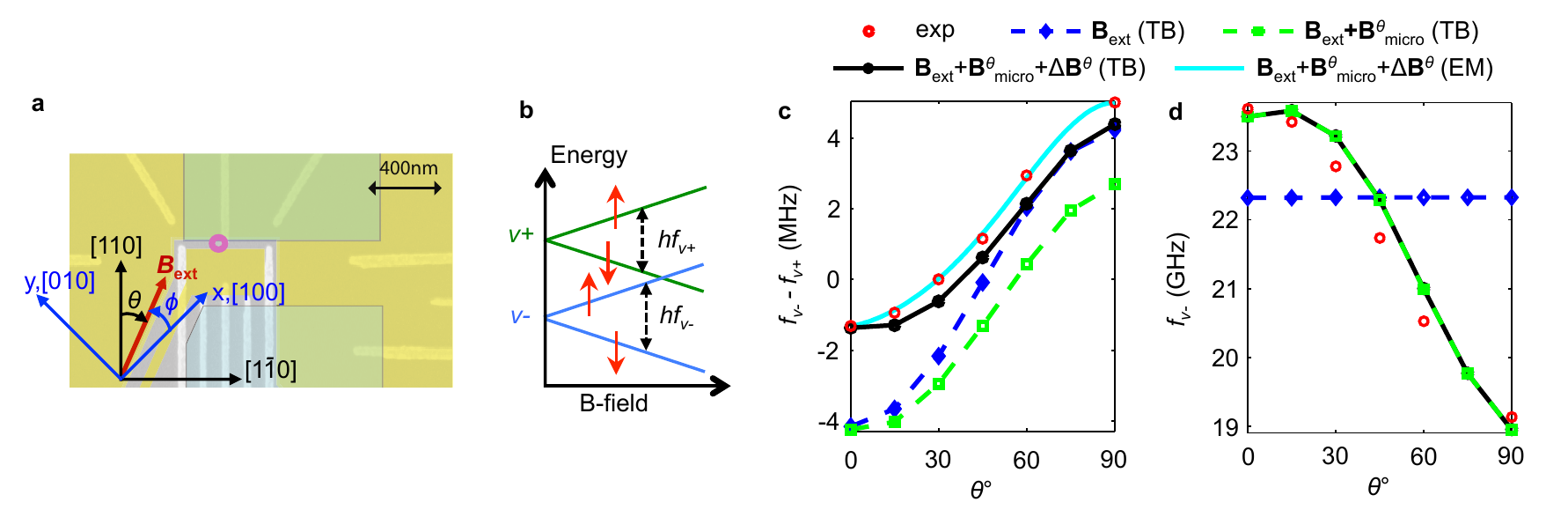}
\caption{{Valley dependent anisotropic ESR in a Si QD with integrated micro-magnets} \textbf{a,} False-color image of the experimental device showing the estimated location of the quantum dot (magenta colored circle) and two Co micro-magnets (green semi-transparent rectangles). The external magnetic field ($\mathbf{B}_{\textup{ext}}$) was rotated clockwise in-plane, from the ${[110]} \left(\theta=0^{\circ}\right)$ crystal orientation towards ${[1\bar{1}0]} \left(\theta=90^{\circ}\right)$. \textbf{b,} Lowest energy levels of a Si QD in an external magnetic field. The valley-split levels $v_-$ and $v_+$ are found to have unequal Zeeman splittings ($E_{\textup{ZS}}(v_{\pm})=h f_{v \pm}$), with ESR frequencies $f_{v_-} \neq f_{v_+}$. In the experiment, all the measured spin splittings are much larger than the valley splitting and are therefore above the anticrossing point of the spin and valley states. \textbf{c,} Both measured (red circles) and calculated $f_{v_-}-f_{v_+}$ as a function of $\theta$, for ${B_{\textup{ext}}}=$0.8 T. The anisotropy in $f_{v_-}-f_{v_+}$ is governed by both internal (intrinsic SOI) and external (micro-magnetic field) factors. The anisotropy due to the intrinsic SOI, calculated from atomistic tight binding method, for a specifically chosen vertical electric field and interface step configuration, is labeled as `$\mathbf{B}_{\textup{ext}}$ (TB)'. The micro-magnetic field is separated into a homogeneous (${\mathbf{B}_{\textup{micro}}^{\theta}}$) and an inhomogeneous (${\Delta \mathbf{B}^{\theta}}$) part. The inclusion of ${\mathbf{B}_{\textup{micro}}^{\theta}}$ in this case (labeled `${\mathbf{B}_{\textup{ext}}+\mathbf{B}_{\textup{micro}}^{\theta}}$ (TB)'), shifts the curve away from the experiment. The addition of ${\Delta \mathbf{B}^{\theta}}$ introduces additional anisotropy (labeled `${\mathbf{B}_{\textup{ext}}+\mathbf{B}_{\textup{micro}}^{\theta}+\Delta \mathbf{B}^{\theta}}$ (TB)') and shifts the curve towards the experiment. An effective-mass calculation, with fitted SOI and dipole coupling parameters, is also presented with a cyan solid line. \textbf{d,} Both measured (red circles) and calculated $f_{v_-}$, as a function of $\theta$, for ${{B}_{\textup{ext}}}=$0.8 T. Calculation with the intrinsic SOI shows negligible change in GHz scale, while the addition of ${\mathbf{B}_{\textup{micro}}^{\theta}}$ results in anisotropy close to the experimental data. ${\Delta \mathbf{B}^{\theta}}$ has negligible effect on $f_{v_-}$. Hence, the anisotropy of $f_{v_-}$ is mainly dictated by the homogeneous part of the micro-magnetic field.}
\vspace{0cm}
\label{fi1}
\end{figure*}

\section*{Introduction}
How microscopic electronic spins in solids are affected by the crystal and interfacial symmetries has been a topic of great interest over the past few decades and has found potential applications in spin-based electronics and computation \cite{SDatta_apl_1990,SAWolf_science_2001,IŽutić_revmodp_2004,DLoss_pra_1998,JRPetta_science_2005,FHLKoppens_nature_2006,RHanson_revmodp_2007}. While the coupling between spin and orbital degrees of freedom has been extensively studied, the interplay between spin and the momentum space valley degree of freedom is a topic of recent interest. 
 This spin-valley interaction is observed in the exotic class of newly found two-dimensional materials \cite{DXiao_prl_2012,ZGong_natcomm_2013,XXu_natphys_2014}, in carbon nanotubes\cite{laird_natnano_2013} and in silicon \cite{VRenard_natcomm_2015,yang_natcomm_2013,XHao_natcomm_2014} \textemdash the old friend of the electronics industry. Progress in silicon qubits in the last few years has come with the demonstrations of various types of qubits with exceptionally long coherence times, such as single spin up/down qubits \cite{veldhrost_natnano_2014,kawakami_natnano_2014}, two-electron singlet-triplet qubits \cite{HRL_2011,wu_pnas_2014}, three-electron exchange-only \cite{HRL_exchange_only} and hybrid spin-charge qubits \cite{kim_nature_2014} and also hole spin qubits\cite{maurand_natcomm_2016} realized in Si QDs. The presence of the valley degree of freedom has enabled valley based qubit proposals\cite{culcer_prl_2012} as well, which have potential for noise immunity. To harness the advantages of different qubit schemes, quantum gates for information encoded in different bases are required\cite{ZGong_natcomm_2013,rohling_njp_2012,rohling_prl_2014}. A controlled coherent interaction between multiple degrees of freedom, like valley and spin, might offer a building block for promising hybrid systems.  

Although bulk silicon has six-fold degenerate conduction band minima, in quantum wells or dots, electric fields and often in-plane strain in addition to vertical confinement  results in only two low lying valley states (labeled as $v_-$ and $v_+$ in Figure 1b) split by an energy gap known as the valley splitting. An interesting interplay between spin and valley degrees of freedom, which gives rise to a valley dependent spin splitting, has been observed in recent experiments\cite{kawakami_natnano_2014,veldhorst_prbrap_2015,Scarlino_prl_2015,Scarlino_arXiv_2016}. SOI enables the control of spin resonance frequencies by gate voltage, an effect measured in refs.\ \citen{veldhrost_natnano_2014,veldhorst_prbrap_2015}. However, the ESR frequencies and their Stark shifts were found to be different for the two valley states \cite{veldhorst_prbrap_2015}. In another work, an inhomogeneous magnetic field, created by integrated micro-magnets in a Si/SiGe quantum dot device, was used to electrically drive ESR\cite{kawakami_natnano_2014}. Magnetic field gradients generated in this way act as an extrinsic spin-orbit coupling and thus can affect the ESR frequency\cite{tokura_prl_2006}. Remarkably, although SOI is a fundamental effect arising from the crystalline structure, the ESR frequency differences between the valley states observed in refs.\ \citen{kawakami_natnano_2014} and \citen{veldhorst_prbrap_2015} have different signs when the external fields are oriented in the same direction with respect to the crystal axes. To understand and achieve control over the coupled behavior between spin and valley degrees of freedom, several questions need to be addressed, such as 1) What causes the device-to-device variability?,  2) Can an artificial source of interaction, like inhomogeneous B-field, completely overpower the SOI effects of the intrinsic material?, 3) What knobs and device designs can be utilized to engineer the valley dependent spin splittings?


\section*{Results}      
Here we report experimentally measured anisotropy in the ESR frequencies of the valley states $f_{v_-}$ and $f_{v_+}$ and their differences $f_{v_-}-f_{v_+}$, as a function of the direction of the external magnetic field ($\mathbf{B}_{\textup{ext}}$) in a quantum dot formed at a Si/SiGe heterostructure with integrated micro-magnets. At specific angles of the external B-field, we also measure the spin splittings of the two valley states as a function of the B-field magnitude. By performing spin-resolved atomistic tight binding (TB) calculations of the quantum dots confined at ideal versus non-ideal interfaces, we evaluate the contribution of the intrinsic SOI with and without the spatially varying B-fields from the micro-magnets to the spin splittings, thereby relating these quantities to the microscopic nature of the interface and elucidating how spin, orbital and valley degrees of freedom are intertwined in these devices. 
Finally, by combining all the effects together, we explain the experimental measurements and address the questions raised in the previous paragraph. 

Fig. 1 shows the experimental device, energy levels of interest and measured anisotropic spin splittings compared with the final theoretical results. Details of the device shown in Fig.\ 1a and the measurement technique of the spin resonance frequency can be found in ref.\ \citen{kawakami_natnano_2014}. 
The external magnetic field is swept from the ${[110]}$ to ${[1\bar{1}0]}$ crystal orientation. A schematic of the energy level structure is shown in Fig.\ 1b depicting the $v_-$ and $v_+$ valley states with different spin splittings, where $v_-$ is defined as the ground state. In the experiment, the lowest valley-orbit excitation is well below the next excitation, justifying this four-level schematic in the energy range of interest. 

\begin{figure}[htbp]
\includegraphics[]{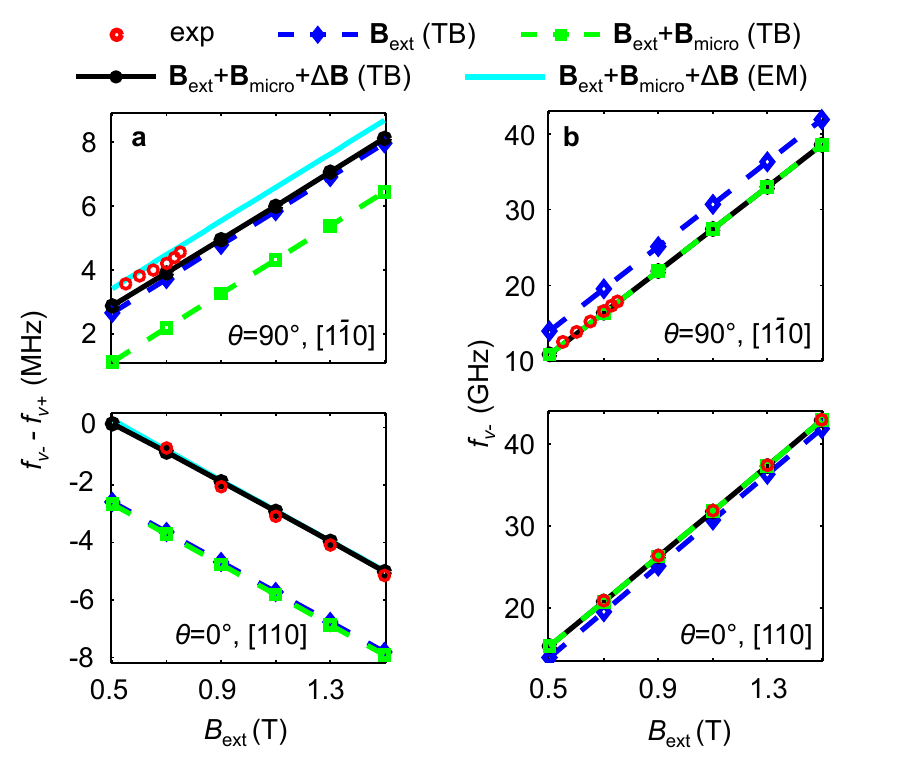}
\caption{{Measured ESR frequencies, ($f_{v_{\pm}}$) and their differences for the two valley states as a function of the external B-field magnitude $B_{\textup{ext}}$ along two crystal directions, and comparison with theoretical calculations.} \textbf{a,} $f_{v_-}-f_{v_+}$ and \textbf{b,} $f_{v_-}$ with $\mathbf{B}_{\textup{ext}} $ along ${[110]} \left(\theta=0^{\circ}\right)$ (bottom panel) and ${[1\bar{1}0]} \left(\theta=90^{\circ}\right)$ (top panel). As in Figs.\ 1c and 1d, the calculations progressively include SOI (labeled `$\mathbf{B}_{\textup{ext}}$ (TB)'), homogeneous (labeled `${\mathbf{B}_{\textup{ext}}+\mathbf{B}_{\textup{micro}}}$ (TB)'), and gradient (labeled `${\mathbf{B}_{\textup{ext}}+\mathbf{B}_{\textup{micro}}+\Delta \mathbf{B}}$ (TB)') B-field of the micro-magnet. The cyan solid lines are the effective mass calculations and the red circles are the experimental data. The dependence (slope) of $f_{v_-}-f_{v_+}$ on $B_{\textup{ext}}$ in (a) comes from the SOI, while the micro-magnetic fields provide a shift independent of $B_{\textup{ext}}$. }  
\vspace{0cm}
\label{fi2}
\end{figure}  

\begin{figure*}[htbp]
\includegraphics[]{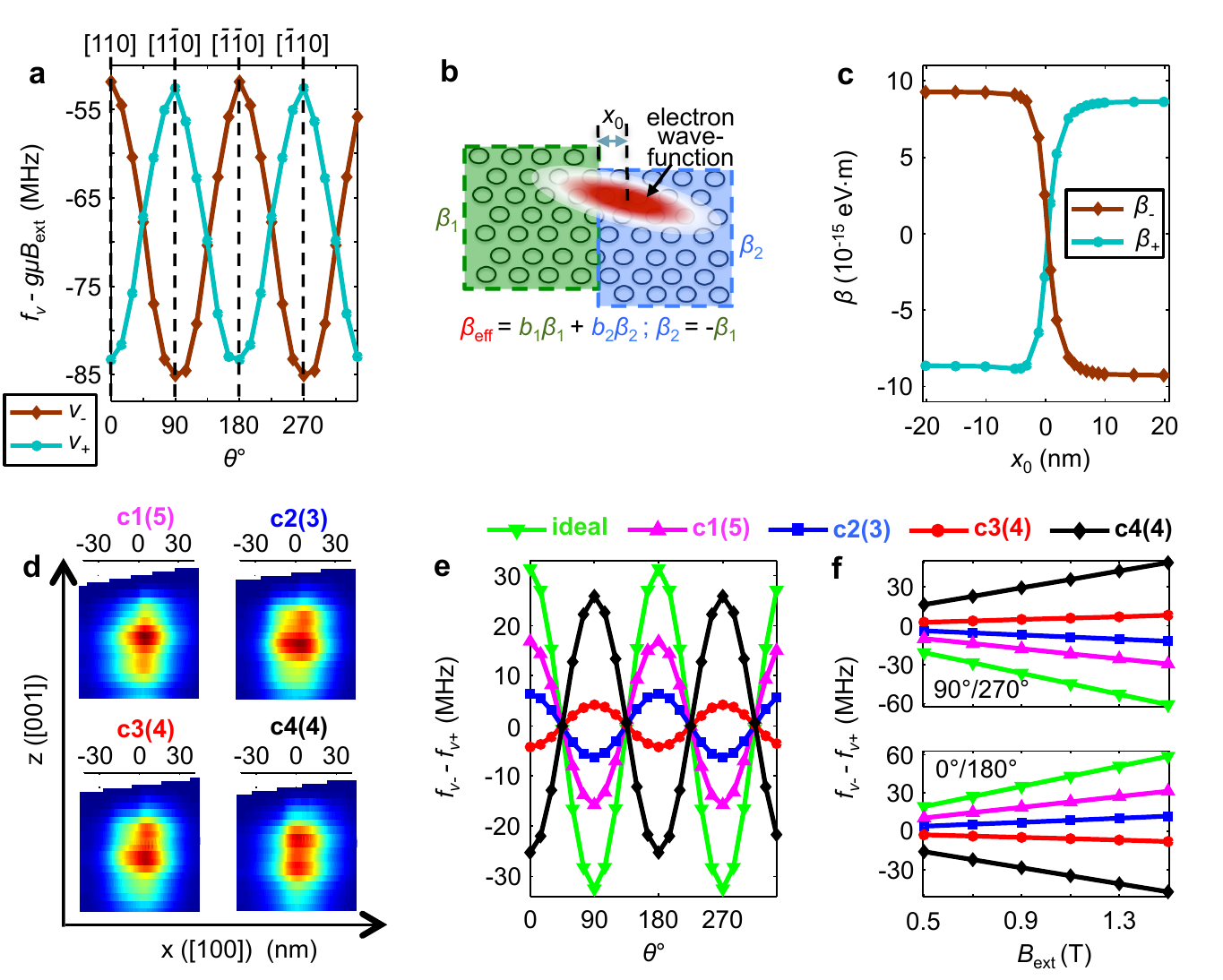}
\caption{{Effect of the intrinsic SOI on ${f_{v_\pm}}$ in a Si QD}. \textbf{a,} Calculated $f_{v \pm}$ as a function of $\theta$, in a QD with ideal (flat) interface, for $B_{\textup{ext}}=$0.8 T, without any micro-magnet. The anisotropies in these curves are in the MHz range and will appear flat on a GHz scale, like the SOI line (labeled $\mathbf{B}_{\textup{ext}}$ (TB)) of Fig.\ 1d. \textbf{b,} Schematic of a QD wavefunction near a monoatomic step at the interface. The distance between the dot center and the step edge is denoted by $x_0$. \textbf{c,} Computed Dresselhaus parameters $\beta_{\pm}$ as a function of $x_0$. $\beta_{\pm}$ changes sign between the two sides of the step. \textbf{d,} QD wave-functions subjected to multiple interface steps. Four different cases are shown (c1(5), c2(3), c3(4), c4(4)) that are used in Figs.\ 3e, 3f and also in Figs.\ 4c and 4d. The number in parentheses is the total number of steps within the QD. Though c3 and c4 has the same number of steps, the location of the steps are different. c3 is the step configuration used in Figs.\ 1 and 2. \textbf{e,} Calculated $f_{v_-}-f_{v_+}$  as a function of $\theta$, for different interface conditions, for $B_{\textup{ext}}=$0.8 T. Interface steps affect both the magnitude and sign of $f_{v_-}-f_{v_+}$. \textbf{f,} $f_{v_-}-f_{v_+}$ with respect to $B_{\textup{ext}}$ along ${[110]}$(${\theta=0^{\circ}}$)/ ${[\bar{1}\bar{1}0]}$ (${\theta=180^{\circ}}$) (bottom panel) and ${[1\bar{1}0]}$ (${\theta=90^{\circ}}$)/${[\bar{1}10]}$ (${\theta=270^{\circ}}$) (top panel). $f_{v_-}-f_{v_+}$ for c3 (red lines with circular marker), in both Figs.\ 3e and 3f, corresponds to the SOI lines (blue dashed lines with diamond marker) of Figs.\ 1c and 2a. The parabolic confinement and $E_z$ used here are the same as that of Figs.\ 1 and 2, except for Fig.\ 3c, where a smaller dot (with a parabolic confinement in both x and y corresponding to orbital energy splitting of 9.4 meV) is used to accommodate for large variation in dot location.}  
\vspace{0cm}
\label{fi3}
\end{figure*}  



The atomistic calculation with SOI alone (labeled `$\mathbf{B}_{\textup{ext}}$ (TB)') for a QD at a specifically chosen non-ideal interface and vertical electric field ($E_z$) qualitatively captures the experimental trend of $f_{v_-}-f_{v_+}$ in Fig.\ 1c, but fails to reproduce the anisotropy of the measured $f_{v_-}$ in Fig.\ 1d in the larger GHz scale. The differences between the experimental data and the SOI-only calculations in both figures arise from the micro-magnets present in the experiment. We can separate the contribution from the micro-magnet into two parts, a homogeneous (spatial average, ${\mathbf{B}_{\textup{micro}}^{\theta}}$) and an inhomogeneous (spatially varying, ${{{\Delta \mathbf{B}}}^{\theta}}$) magnetic field. The superscript $\theta$ here indicates that the micro-magnetic fields depend on the direction of $\mathbf{B}_{\textup{ext}}$ (supplementary section S4). The inclusion of the homogeneous part of the micro-magnetic field creates an anisotropy in the total magnetic field (supplementary Fig.\ S7), which captures the anisotropy of $f_{v_-}$ 
in Fig.\ 1d very well ($f_{v_-}\approx g\mu  \left| {{\mathbf{B}_{\textup{ext}}}}+{\mathbf{B}_{\textup{micro}}^{\theta }} \right| /h$, where $g$ is the Land\'e g-factor, $\mu$ is the Bohr magneton and $h$ is the Planck constant), but quantitative match with the experimental data in Fig.\ 1c is not obtained. Next, we also incorporate the inhomogeneous part of the micro-magnetic field, and witness a close quantitative agreement in the anisotropy of $f_{v_-}-f_{v_+}$, 
while the anisotropy of $f_{v_-}$ is unaffected. This experiment-theory agreement of fig.\ 1c is achieved for a specific choice of interface condition and $E_z$, whose influence will be discussed later. Here, we conclude that mainly the intrinsic SOI and the extrinsic inhomogeneous B-field govern the anisotropy of $f_{v_-}-f_{v_+}$ on the MHz scale, while the anisotropy in the total homogeneous magnetic field introduced by the micro-magnet dictates the anisotropy of $f_{v_-}$ (and $f_{v_+}$) on the larger GHz scale.

In Fig.\ 2, we show the measurements of the spin splittings 
 as a function of the magnitude of $\mathbf{B}_{\textup{ext}}$ ($B_{\textup{ext}}$), together with the theoretical calculations. The bottom panels show $f_{v_-}-f_{v_+}$ (Fig. 2a)  and $f_{v_-}$ (Fig. 2b) for $\mathbf{B}_{\textup{ext}}$ along ${[110]}$ ($\theta=0{^\circ}$), whereas the top panels correspond to the B-field along ${[1\bar{1}0]}$ ($\theta=90^{\circ}$). In Fig. 2b, $f_{v_-}$ depends on $B_{\textup{ext}}$ through $g_- \mu B_{\textup{tot}}/h$, with $B_{\textup{tot}}=|\mathbf{B}_{\textup{ext}}+{\mathbf{B}_{\textup{micro}}}|$. The addition of ${\mathbf{B}_{\textup{micro}}}$ causes a change in $B_{\textup{tot}}$ and shifts $f_{v_-}$ to coincide with the experimental data. The contributions of ${\Delta \mathbf{B}}$ and SOI are negligible here in the GHz scale.

On the other hand, comparing the calculated $f_{v_-}-f_{v_+}$ from SOI alone (labeled `$\mathbf{B}_{\textup{ext}}$ (TB)') for the chosen $E_z$ and interface condition, with experimental data, in both the top and bottom panels of Fig.\ 2a, it is clear that the experimental B-field dependence of $f_{v_-}-f_{v_+}$ (the slope, $\frac{d(f_{v_-}-f_{v_+})}{dB_{\textup{ext}}}$) is captured from the effect of intrinsic SOI, except for a shift between the SOI curve and the experimental data (different shift for $\theta=0^\circ$ and $\theta=90^\circ$). The addition of ${\mathbf{B}_{\textup{micro}}}$ alone does not result in the necessary shift to match the experiment. Only after adding ${\Delta \mathbf{B}}$ can a quantitative match with the experiment be achieved. Again the experiment-theory agreement is conditional on the interface condition and $E_z$. Moreover, we see that the addition of ${\Delta \mathbf{B}}$ does not change the dependency on $B_{\textup{ext}}$. Therefore, to properly explain the observed experimental behavior, we can ignore neither the SOI, which is responsible for the change in $f_{v_-}-f_{v_+}$ with $B_{\textup{ext}}$, nor the inhomogeneous B-field which shifts $f_{v_-}-f_{v_+}$ regardless of $B_{\textup{ext}}$.

\section*{Discussion}
To obtain a quantitative agreement between the experiment and the atomistic TB calculations, simultaneously in the anisotropy (Fig.\ 1c) and the $B_{\textup{ext}}$ (Fig.\ 2a) dependence of $f_{v_{-}}-f_{v_{+}}$, the only knobs we have to adjust are \begin{enumerate*} \item $E_z$ and \item interfacial geometry \end{enumerate*} i.e. how many atomic steps at the interface lie inside the dot and where they are located relative to the dot center. These adjustments have to be done iteratively since the steps and $E_z$ not only affect the intrinsic SOI but also the influence of the inhomogeneous B-field. It is easy to separate out the contribution of the SOI from the micro-magnet in the $B_{\textup{ext}}$ dependence of $f_{v_{-}}-f_{v_{+}}$. It will be shown in Figs.\ 3 and 4 that, the slope, $\frac{d(f_{v_-}-f_{v_+})}{dB_{\textup{ext}}}$ originates from the SOI, while the micro-magnetic field shifts $f_{v_{-}}-f_{v_{+}}$ independent of $B_{\textup{ext}}$. First we individually match the experimental ``slope" from the SOI and the ``shift" from the contribution of the micro-magnet for some combinations of the two knobs. Finally both effects together quantitatively match the experiment for $E_z=6.77$ MVm$^{-1}$, 
and an interface with four evenly spaced monoatomic steps at -24.7 nm, -2.9 nm, 18.7 nm, 40.4 nm from the dot center along the x (${[100]}$) direction. This combination also predicts a valley splitting of 34.4 $\mu$eV in close agreement with the experimental value, given by 29 $\mu$eV \cite{Scarlino_arXiv_2016}. To describe the QD, a 2D simple harmonic (parabolic confinement) potential was used with orbital energy splittings of 0.55 meV and 9.4 meV characterizing the x and y (${[010]}$) confinement respectively. As the interface steps are parallel to y direction, the orbital energy splitting along y has negligible effects, but the strong y confinement significantly reduces simulation time. 

To further our understanding, we have complemented the atomistic calculations with an effective mass (EM) based analytic model with Rashba and Dresselhaus-like SOI terms [supplementary section S1], as used in earlier works \cite{nestoklon_prb_2006, nestoklon_prb_2008, veldhorst_prbrap_2015,golub_prb_2004}. We have also developed an analytic model to capture the effects of the inhomogeneous magnetic field [supplementary section S2]. The contributions of the SOI and ${\Delta \mathbf{B}}$ on $f_{v_-}-f_{v_+}$ obtained from these models are shown in equations \ref{equ1} and \ref{equ2} respectively.



\begin{multline}
\Delta \left(f_{{{v}_{-}}}-f_{{{v}_{+}}}\right)^{\textup{SOI}} \approx \\
\frac{4\pi \left| e \right|\left\langle z \right\rangle }{{{h}^{2}}}{B_{\textup{ext}}}\Big\{ \left( {{\beta }_{-}}-{{\beta }_{+}} \right)\sin 2\phi -\left( {{\alpha }_{-}}-{{\alpha }_{-}} \right) \Big\}
\label{equ1}
\end{multline}

\begin{multline}
\Delta \left(f_{{{v}_{-}}}-f_{{{v}_{+}}}\right)^{\Delta \mathbf{B}} \approx \\ 
\frac{g\mu }{h}\Bigg\{\cos \phi \left( \left( \left\langle {{x}_{-}} \right\rangle -\left\langle {{x}_{+}} \right\rangle  \right)\frac{dB_{x}^{\phi }}{dx}+\left( \left\langle {{y}_{-}} \right\rangle -\left\langle {{y}_{+}} \right\rangle  \right)\frac{dB_{x}^{\phi }}{dy} \right) \\
+\sin \phi \left( \left( \left\langle {{x}_{-}} \right\rangle -\left\langle {{x}_{+}} \right\rangle  \right)\frac{dB_{y}^{\phi }}{dx}+\left( \left\langle {{y}_{-}} \right\rangle -\left\langle {{y}_{+}} \right\rangle  \right)\frac{dB_{y}^{\phi }}{dy} \right)\Bigg\} 
\label{equ2}
\end{multline}

\noindent Here, $\alpha_{\pm}$ and $\beta_{\pm}$ are the Rashba and Dresselhaus-like coefficients respectively, $\left\langle z \right\rangle$ is the spread of the electron wavefunction along z, $\left\langle{x}_{\pm }\right\rangle$ and $\left\langle{y}_{\pm }\right\rangle$ are the intra-valley dipole matrix elements, $\phi$ is the angle of the external magnetic field with respect to the ${[100]}$ crystal orientation and $\frac{dB_i^{\phi}}{dj}$ are the magnetic field gradients along different directions ($i,j=x,y,z$) for a specific angle $\phi$. It is clear from these expressions that to match $f_{v_-}-f_{v_+}$ the difference in SOI and dipole moment parameters between the valley states are relevant (but not their absolute values). The parameters used to match the experiment are $\alpha_- -\alpha_+=-2.5370\times 10^{-15}$ eV$\cdot$m, $\beta_- -\beta_+=9.4564\times 10^{-19}$ eV$\cdot$m, $\left\langle {{x}_{-}} \right\rangle -\left\langle {{x}_{+}} \right\rangle=-0.169$ nm, $\left\langle {{y}_{-}} \right\rangle -\left\langle {{y}_{+}} \right\rangle=0$ nm and $\left\langle z \right\rangle=2.792$ nm. These fitting parameters in the EM calculations enable us to obtain an even better match with the experimental data compared to TB in Figs.\ 1c and 2a (cyan solid lines). Here we want to point out that the accurateness of the numerically calculated micro-magnetic field values depends on our estimation of the dot location. But as we calculate $\left( {{\beta }_{-}}-{{\beta }_{+}} \right)$ and $\left( {{\alpha }_{-}}-{{\alpha }_{-}} \right)$ independently by comparing the measured $\frac{d(f_{v_-}-f_{v_+})}{dB_{\textup{ext}}}$ for [110] and ${[1\bar{1}0]}$ with equation \ref{equ1}(supplementary section S5), any uncertainty in the estimated dot location or the micro-magnetic field values does not effect the extracted SOI parameters. 



As shown in Figs.\ 1 and 2, three physical attributes play a key role in explaining the experimental data, 1) SOI, 2) ${\mathbf{B}_{\textup{micro}}}$, and 3) ${\Delta \mathbf{B}}$. Each of these contribute to $f_{v_{\pm}}$, and only their sum can accurately reproduce the experimental data for a specific interface condition and vertical electric field, the two knobs mentioned in earlier paragraph. In Figs.\ 3 and 4, we show separately the effects of 1) and 3) respectively. We show how the contributions of SOI and ${\Delta \mathbf{B}}$ are modulated by the nature of the confining interface (knob 2). The influence of $E_z$ (knob 1) on the effects of SOI and ${\Delta \mathbf{B}}$ are shown in the supplementary Figs.\ S3 and S4 respectively. We also show how ${\mathbf{B}_{\textup{micro}}}$ modifies the total homogeneous B-field in the supplementary Fig.\ S7.   



Fig.\ 3a shows the angular dependence of $f_{v_{\pm}}$ for a Si QD with a smooth interface calculated from TB. Both $f_{v_-}$ and $f_{v_+}$ show a 180$^{\circ}$ periodicity but they are 90$^{\circ}$ out of phase. From analytic effective mass study [supplementary equation S11], we understand that the anisotropic contribution from the Dresselhaus-like interaction, caused by interface inversion asymmetry\cite{nestoklon_prb_2008}, results in this angular dependence in $f_{v_{\pm}}$. Moreover, the different signs of the Dresselhaus coefficients $\beta_{\pm}$ for the valley states, give rise to a 90$^{\circ}$ phase shift between $f_{v_-}$ and $f_{v_+}$. It is important to notice that the change in $f_{v_{\pm}}$ is in MHz range. So, in GHz scale, like the blue curve (diamond markers) in Fig.\ 1 d, this change is not visible. However, if we compare $f_{v_-}$ and $f_{v_+}$ for this ideal interface case, we see $f_{v_-}>f_{v_+}$ at $\theta=$0$^{\circ}$ and $f_{v_-}<f_{v_+}$ at $\theta=$90$^{\circ}$, which does not explain the experimentally measured anisotropy. We now discuss the remaining physical parameters needed to obtain a complete understanding of the experiment.

It is well-known that the interface between Si/SiGe or Si/SiO$_2$ has atomic-scale disorder, with monolayer atomic steps being a common form of disorder \cite{steps_justify}. To understand how such non-ideal interfaces can affect SOI, we first introduce a monolayer atomic step as shown in Fig.\ 3b and vary the dot position laterally relative to the step, as defined by the variable $x_0$. By fitting the EM solutions to the TB results [supplementary equation S15], we have extracted the Dresselhaus-like coefficient $\beta_{\pm}$ and plotted it in Fig.\ 3c as a function of $x_0$. It is seen that $\beta_{\pm}$ changes sign as the dot moves from the left to the right of the step edge. Both the sign and magnitude of $\beta_{\pm}$ depends on the distribution of the wavefunction between the neighboring regions with one atomic layer shift between them, as shown in fig.\ 3b. A monoatomic shift of the vertical position of the interface results in a 90$^{\circ}$ rotation of its atomic arrangements about the [001] axis, which results in a sign inversion of the Dresselhaus coefficient of that region\cite{nestoklon_prb_2008}. A dot wavefunction spread over a monoatomic step therefore samples out a weighted average of two $\beta$s with opposite signs\cite{golub_prb_2004,nestoklon_prb_2006}. 


\begin{figure}[htbp]
\includegraphics[trim=0 0 0 0, clip, width=\linewidth]{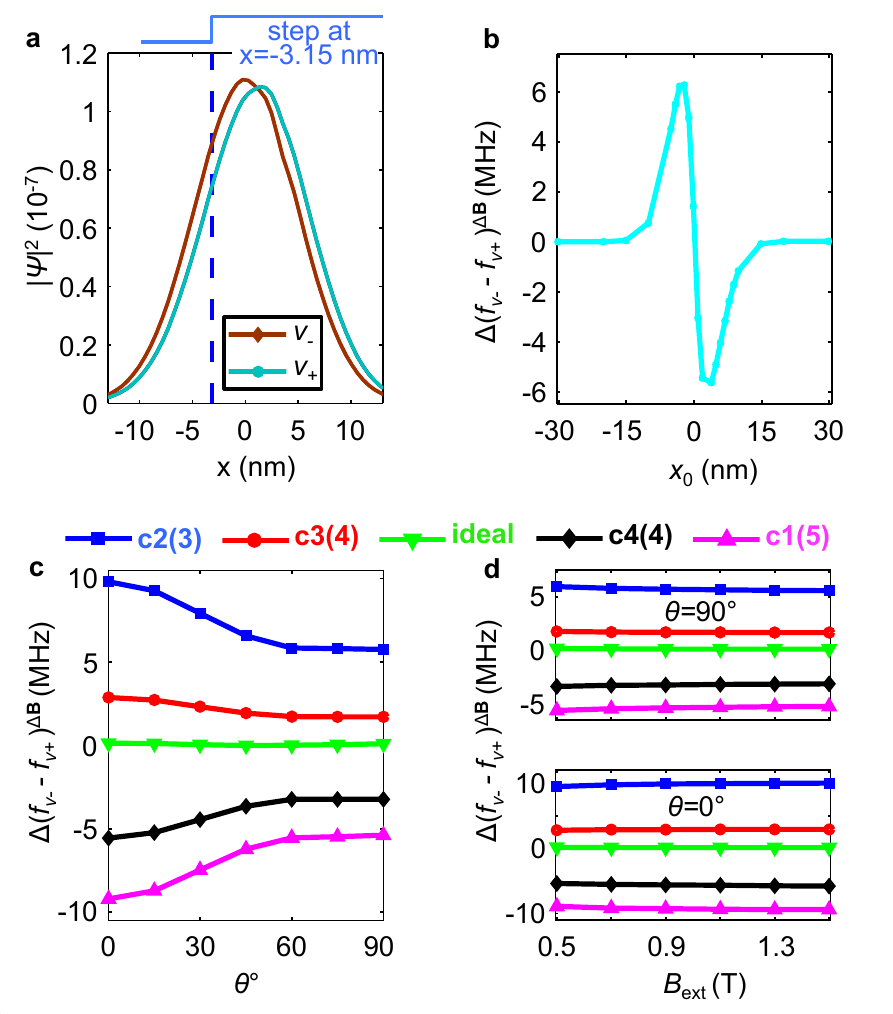}
\caption{{Effect of inhomogeneous magnetic field on ${f_{v_{-}}-f_{v_+}}$.} \textbf{a,} 1D cut of the wavefunctions of the two valley states close to a step edge, highlighting their spatial differences. A large vertical E-field, $E_z=$30 MVm$^{-1}$ is used here to show a magnified effect. \textbf{b,} The change in $f_v{_-}-f_v{_+}$ due to the inhomogeneous B-field (${\Delta \mathbf{B}}$) alone as a function of the distance $x_0$ between the dot center and a step edge, as defined in Fig. 3b. \textbf{c,} Angular dependence of $\Delta (f_v{_-}-f_v{_+})^{{\Delta \mathbf{B}}}$ for the various step configurations of Fig. 3d (same color code) computed from atomistic TB. \textbf{d,} $\Delta (f_v{_-}-f_v{_+})^{{\Delta \mathbf{B}}}$ as a function of $B_{\textup{ext}}$. $\Delta (f_v{_-}-f_v{_+})^{{\Delta \mathbf{B}}}$ shows negligible dependence on $B_{\textup{ext}}$. $\Delta (f_v{_-}-f_v{_+})^{{\Delta \mathbf{B}}}$ for c3 (red lines with circular marker), in Figs.\ 4c and 4d, corresponds to the contribution of ${\Delta \mathbf{B}}$ (the difference between the black solid curve/lines with circular markers and green dashed curve/lines with square markers)  of Figs.\ 1c and 2a. The fields used in the simulations of c and d are the same as that of Figs.\ 1 and 2, whereas the fields used for b are the same as that of Fig.\ 3c.}  
\vspace{0cm}
\label{fi4}
\end{figure}  

Next, we investigate the anisotropy of $f_{v_-}-f_v{_+}$ (Fig.\ 3e) with various step configurations shown in Fig.\ 3d. $f_{v_-}-f_v{_+}$ in Fig.\ 3e exhibits a 180$^{\circ}$ periodicity, with extrema at the ${[110]}$, ${[1\bar{1}0]}$, ${[\bar{1}\bar{1}0]}$, ${[\bar{1}10]}$ crystal orientations. Both the sign and magnitude of $f_{v_-}-f_v{_+}$ depends on the interface condition. Since $\beta_{\pm}$ decreases when a QD wavefunction is spread over a step edge, the smooth interface case (green curve) has the highest amplitude. Fig.\ 3f shows that the slope of $f_{v_-}-f_v{_+}$ with $B_{\textup{ext}}$ changes sign for a 90$^{\circ}$ rotation of $\mathbf{B}_{\textup{ext}}$ and is strongly dependent on the step configuration. The step configuration labeled c3 in Fig.\ 3d is used to match the experiment in Figs.\ 1 and 2. So the curves for c3 in both Figs.\ 3e and 3f correspond to the SOI results of Figs.\ 1c and 2a. It is key to note here that, as $E_z$ also influences $\left|f_{v_-}-f_{v_+}\right|$ and $\left|\frac{d(f_{v_-}-f_{v_+})}{dB_{\textup{ext}}}\right|$, shown in supplementary Fig.\ S3, a different combination of interface steps and $E_z$ can also produce these same SOI results of Figs.\ 1c and 2a, but might not result in the necessary contribution from micro-magnet to match the experiment. Now the dependence of $f_{v_-}-f_{v_+}$ on the interface condition will cause device-to-device variability, while the dependence on the direction and magnitude of $\mathbf{B}_{\textup{ext}}$ can provide control over the difference in spin splittings. These results thus give us answers to the questions 1 and 3 asked in paragraph 3. 




Fig.\ 4 illustrates how the inhomogeneous magnetic field alone changes $f_{v_-}-f_{v_+}$ (denoted as $\Delta (f_{v_-}-f_{v_+})^{{\Delta \mathbf{B}}}$). Since ${\Delta \mathbf{B}^{\theta}}$ vectorially adds to $\mathbf{B}_{\textup{ext}}$, an anisotropic $\Delta (f_{v_-}-f_{v_+})^{{\Delta \mathbf{B}}}$ is seen in Fig.\ 4c with and without the various step configurations portrayed in Fig.\ 3d. We also see that $\Delta (f_{v_-}-f_{v_+})^{{\Delta \mathbf{B}}}$ in Fig. 4c is negligible for a flat interface, but is significant when interface steps are present. This can be understood from Figs.\ 4a and 4b, and/or equation \ref{equ2}. Interface steps generate strong valley-orbit hybridization\cite{gamble_prb_2013,Friesen_steps} causing the valley states to have non-identical wavefunctions, and hence different dipole moments, $\left( \left\langle {{x}_{-}} \right\rangle -\left\langle {{x}_{+}} \right\rangle  \right)\ne 0$ and/or  $\left( \left\langle {{y}_{-}} \right\rangle -\left\langle {{y}_{+}} \right\rangle  \right)\ne 0$, as opposed to a flat interface case, which has  $\left\langle {{x}_{\pm }} \right\rangle=\left\langle {{y}_{\pm }} \right\rangle=0$. Thus the spatially varying magnetic field has a different effect on the two wavefunctions, thereby contributing to the difference in ESR frequencies between the valley states. Fig.\ 4b shows $\Delta (f_{v_-}-f_{v_+})^{{\Delta \mathbf{B}}}$ as a function of the dot location relative to a step edge, $x_0$ (as in Fig.\ 3b) and illustrates that ${\Delta \mathbf{B}}$ has the largest contribution to $f_{v_-}-f_{v_+}$ when the step is in the vicinity of the dot. Also, $\Delta (f_{v_-}-f_{v_+})^{{\Delta \mathbf{B}}}$ is almost independent of $B_{\textup{ext}}$, as shown in Fig.\ 4d. The curves labeled c3 in both Figs.\ 4c and 4d correspond to the contribution of ${\Delta \mathbf{B}}$ in Figs.\ 1c and 2a. Now $E_z$ also influences $\left|\Delta (f_{v_-}-f_{v_+})^{{\Delta \mathbf{B}}}\right|$, as shown in supplementary Fig.\ S4. Thus a different combination of interface steps and $E_z$ can also produce these same ${\Delta \mathbf{B}}$ results of Figs.\ 1c and 2a, but might not result in the necessary SOI contribution to match the experiment.


A comparison between Figs.\ 3f and 4d (also between equations \ref{equ1} and \ref{equ2}) reveals that any dependence of $f_{v_-} -f_{v_+}$ on $B_{\textup{ext}}$ can only come from the SOI. This indicates that the experimental B-field dependency in Fig.\ 2a can not be explained without the SOI. So the effect of the SOI cannot be ignored even in the presence of a micro-magnet and this answers question 2 raised in the third paragraph. However, engineering the micro-magnetic field will allow us to engineer the anisotropy of $f_{v_-}-f_{v_+}$ (question 3, paragraph 3). Also, the influence of interface steps will cause additional device-to-device variability (question 1, paragraph 3). 

\begin{figure}[htbp]
\includegraphics[]{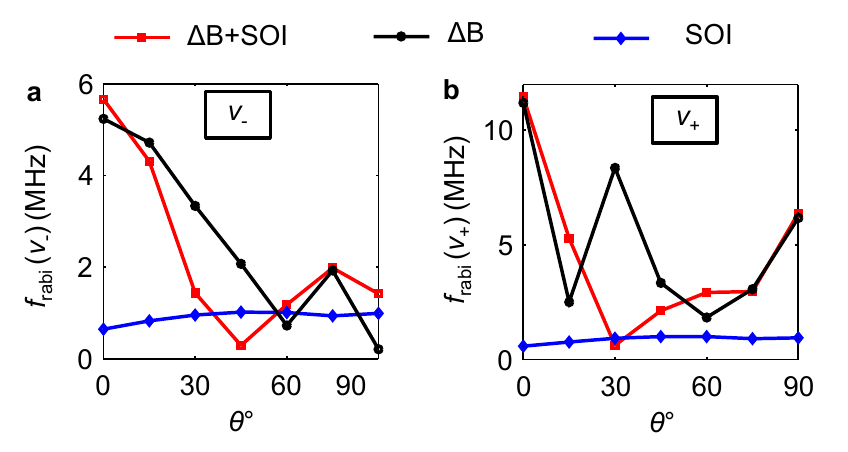}
\caption{Calculated Rabi frequencies ($f_{\textup{Rabi}}$) with SOI only, inhomogeneous B-field only and both SOI and inhomogeneous B-field for different direction of the external magnetic field for both $v_-$(panel \textbf{A}) and $v_+$(panel \textbf{B}) valley states. Interface condition, vertical e-field and parabolic confinement for the dot used in these simulations are the same as that used to match the experimental data in Figs.\ 1 and 2. The valley and orbital splittings that we get from the simulations are around 34.4 $\mu$eV and 0.48 meV respectively. The dot radius is around 35 nm. The fastest Rabi frequencies for SOI only are around 1MHz, which are least five times smaller compared to that of the gradient B-field for $\theta=0^{\circ}$. It is important to note here that the micro-magnet geometry was designed to maximize the Rabi frequency at $\theta=0^{\circ}$. The details of the $f_{\textup{Rabi}}$ calculation are discussed in supplementary section S6.}
\vspace{0cm}
\label{fi8}
\end{figure}

Now the understanding of an enhanced SOI effect compared to bulk, brings forward an important question, whether it is possible to perform electric-dipole spin resonance (EDSR) without the requirement of micro-magnets. Here, we predict that (Fig.\ 5) for similar driving amplitudes as used here the SOI-only EDSR can offer Rabi frequencies close to 1 MHz, which is around five times smaller than the micro-magnet based EDSR. Moreover, the Rabi frequency of the SOI-EDSR will strongly depend on the interface condition\cite{wister_prb_2017} (supplementary section S6) and can be difficult to control or improve. On the other hand, with improved design (stronger transverse gradient field)  we can gain more advantage of the micro-magnets and drive even faster Rabi oscillations. However, we also predict that, both the SOI and inhomogeneous B-field contribute to the $E_z$ dependence of $f_{v_{\pm}}$ (supplementary section S3) and make the qubits susceptible to charge noise\cite{veldhorst_prbrap_2015}. As these two have comparable contribution, both of their effects will add to the charge noise induced dephasing of the spin qubits in the presence of micro-magnets. 

The coupled spin and valley behavior observed in this work may in principle enable us to simultaneously use the quantum information stored in both spin and valley degrees of freedom of a single electron. For example, a valley controlled not gate\cite{ZGong_natcomm_2013} can be designed in which the spin basis can be the target qubit, while the valley information can work as a control qubit. If we choose such a direction of the external magnetic field, where the valley states have different spin splittings, an applied microwave pulse in resonance with the spin splitting of $v_{-}$, will rotate the spin only if the electron is in $v_{-}$. So we get a NOT operation of the spin quantum information controlled by the valley quantum information. Spin transitions conditional to valley degrees of freedom are also shown in ref.\ \citen{kawakami_natnano_2014} and an inter-valley spin transition, which can entangle spin and valley degrees of freedom, is observed in ref.\ \citen{Scarlino_arXiv_2016}. 

\section*{Conclusion}
To conclude, we experimentally observe anisotropic behavior in the electron spin resonance frequencies for different valley states in a Si QD with integrated micro-magnets. We analyze this behavior theoretically and find that intrinsic SOI introduces 180$^\circ$ periodicity in the difference in the ESR frequencies between the valley states, but the inhomogeneous B-field of the micro-magnet also modifies this anisotropy. Interfacial non-idealities like steps control both the sign and magnitude of this difference through both SOI and inhomogeneous B-field. We also measure the external magnetic field dependence of the resonance frequencies. We show that the measured magnetic field dependence of the difference in resonance frequencies originates only from the SOI. We conclude that even though the SOI in bulk silicon has been typically ignored as being small, it still plays a major role in determining the valley dependent spin properties in interfacially confined Si QDs. These understandings help us answer the questions raised in paragraph 3, which are crucial for proper operation of various qubit schemes based on silicon quantum dots.


\section*{Methods}
For the theoretical calculations, we used a large scale atomistic tight binding approach with spin resolved sp$^3$d$^5$s* atomic orbitals with nearest neighbor interactions\cite{klimeck_ted_2007}. 
Typical simulation domains comprise of 1.5-2 million atoms to capture realistic sized dots. Spin-orbit interactions are directly included in the Hamiltonian as a matrix element between p-orbitals following the prescription of Chadi \cite{chadi_prb_1977}. The advantage of this approach is that no additional fitting parameters are needed to capture various types of SOI such as Rashba and Dresselhaus SOI in contrast to k.p theory. We introduce monoatomic steps as a source of non-ideality consistent with other works\cite{steps_justify, Friesen_steps, kharche_apl_2007}. The Si interface was modeled with Hydrogen passivation, without using SiGe. This interface model is sufficient to capture the SOI effects of a Si/SiGe interface discussed in refs.\ \citen{nestoklon_prb_2008,nestoklon_prb_2006,golub_prb_2004}. 
We have used the methodology of ref.\ \citen{Goldman_apa_2000} to model the micro-magnetic fields [supplementary section S4]. Full magnetization of the micro-magnet is assumed. This causes the value of the magnetization of the micro-magnet to be saturated and makes it independent of $B_{\textup{ext}}$. However, a change in the direction of $B_{\textup{ext}}$ changes the magnetization. We include the effect of inhomogeneous magnetic field perturbatively, with the perturbation matrix elements, $\langle {{\psi }_{m}}|\frac{1}{2}g\mu {\Delta \mathbf{B}^{\phi}}\left| {{\psi }_{n}} \right\rangle =\frac{1}{2}g\mu \sum\limits_{i,j}{\langle {{\psi }_{m}}|\frac{dB_i^{\phi}}{dj}j\left| {{\psi }_{n}} \right\rangle }$. Here, $\psi_n$ and $\psi_m$ are atomistic wave-functions calculated with homogeneous magnetic field. For further details about the numerical techniques, see NEMO3D reference \citen{klimeck_ted_2007}. Method details about the experiment can be found in ref.\ \citen{kawakami_natnano_2014}. The dot location in this experiment is different from ref.\ \citen{kawakami_natnano_2014}. The device was electrostatically reset by shining light using an LED and all the measurements were done with a new electrostatic environment (a new gate voltage configuration). The quantum dot location is estimated by the offsets of the magnetic field created by the micro-magnets extrapolated from the measurements shown in Fig.\ 2 and comparing to the simulation results shown in supplementary section S4. We also observed that the Rabi frequencies were different from ref.\ \citen{kawakami_natnano_2014} when applying the same microwave power to the same gate, which qualitatively indicates that the dot location is different.

\section*{Acknowledgments}
This work was supported in part by ARO (W911NF-12-0607); development and maintenance of the growth facilities used for fabricating samples is supported by DOE (DE-FG02-03ER46028). This research utilized NSF-supported shared facilities (MRSEC
DMR-1121288) at the University of Wisconsin-Madison. Computational resources on nanoHUB.org, funded by the NSF grant EEC-0228390, were used. M.P.N. acknowledges support from ERC Synergy Grant. R.F. and R.R. acknowledge discussions with R. Ruskov, C. Tahan, and A. Dzurak.

\vspace{0.5cm}
\section*{Author contributions statement}
R.F. performed the g-factor calculations, explained the underlying physics and developed the theory with guidance from R.R. R.F., R.R., E.K., P.S., and M.P.N. analyzed the simulation results and compared with experimental data in consultation with L.M.K.V., M.F., S.N.C. and M.A.E.. E.K. and P.S. performed the experiment and analyzed the measured data. D.R.W. fabricated the sample. D.E.S. and M.G.L. grew the heterostructure. R.F. and R.R. wrote the manuscript with feedback from all the authors. R.R. and L.M.K.V. initiated the project, and supervised the work with S.N.C, M.F. and M.A.E.

\section*{Additional information}
The authors declare that they have no competing financial interests.

\end{document}


\section*{Supplementary Materials for `Valley dependent anisotropic spin splitting in silicon quantum dots'}

\date{\today}

\textbf{S1. Analytic effective mass model to explain the effect of spin-orbit interaction (SOI) on the valley dependent g-factor anisotropy in a silicon quantum dot}

To explain our atomistic tight-binding (TB) results of the valley dependent g-factor and its anisotropy in a silicon quantum dot (QD), we have also performed analytic effective mass calculations as shown here. The electron Hamiltonian can be written as,
 
\begin{align}
H=\frac{{{\hbar }^{2}}}{2m}{{\mathbf{k}}^{2}}+V(\mathbf{r})+{{H}_{\textup{Z}}}+{{H}_{\textup{SO}}}
\label{equ1}
\end{align} 

\noindent
Here, $m$ is the electron effective mass, which assumes a value of $0.19m_0$ and $0.91m_0$ for the transverse (x or [100], y or [010]) and longitudinal (z or [001]) components in Si respectively ($m_0$ being the free electron mass). $V(\mathbf{r})$ is the potential defining the quantum dot, and has been discussed in the main text. ${{H}_{Z}}=\frac{1}{2}g\mu \bm{\sigma }\cdot\mathbf{B}$ is the Zeeman term, with $g$ the electron $g$-factor, $\mu$ the Bohr magneton, and $\mathbf{B}$ the applied magnetic field at the dot location. ${{H}_{\textup{SO}}}=\beta \left( {{\sigma }_{x}}{{k}_{x}}-{{\sigma }_{y}}{{k}_{y}} \right)+\alpha \left( {{\sigma }_{x}}{{k}_{y}}-{{\sigma }_{y}}{{k}_{x}} \right)$ is the SOI term, where $\beta$ ($\alpha$) is the strength of Dresselhaus (Rashba) interaction, $\sigma_x$, $\sigma_y$ are the Pauli spin matrices, and $k_x$ ($k_y$) is electron canonical momentum along x (y) direction. 
$\mathbf{k}=-i\bm{\nabla }-e\frac{{\mathbf{A}}}{\hbar }$, where $\mathbf{A}$ is the vector potential and $\mathbf{B}=\bm{\nabla }\times \mathbf{A}$. Now, we treat ${{H}_{\textup{P}}}=\frac{1}{2}g\mu \bm{\sigma }\cdot \mathbf{B}+\beta \left( {{\sigma }_{x}}{{k}_{x}}-{{\sigma }_{y}}{{k}_{y}} \right)+\alpha \left( {{\sigma }_{x}}{{k}_{y}}-{{\sigma }_{y}}{{k}_{x}} \right)$ as a perturbation to ${{H}_{\textup{0}}}=\frac{{{\hbar }^{2}}}{2m}{{\mathbf{k}}^{2}}+V(\mathbf{r})$. The unperturbed Hamiltonian ${{H}_{0}}$ yields the spin degenerate eigenstates of a Si QD. Typically, in these devices, orbital splitting ($E_{{\textup{OS}}}$) is much larger than valley ($E_{\textup{VS}}$) and spin ($E_{\textup{{\textup{ZS}}}}$) splittings. So, we only consider the four lowest energy states $\left| \psi _{{{v}_{-}}}^{\downarrow } \right\rangle $, $\left| \psi _{{{v}_{-}}}^{\uparrow } \right\rangle $, $\left| \psi _{{{v}_{+}}}^{\downarrow } \right\rangle $, $\left| \psi _{{{v}_{+}}}^{\uparrow } \right\rangle $ as the basis for the perturbation calculation.
We can write the perturbation Hamiltonian as,

\begin{align}
{{H}_{\textup{P}}}={{\sigma }_{x}}\left( \frac{1}{2}g\mu {{B}_{x}}+\beta {{k}_{x}}+\alpha {{k}_{y}} \right)+{{\sigma }_{y}}\left( \frac{1}{2}g\mu {{B}_{y}}-\beta {{k}_{y}}-\alpha {{k}_{x}} \right)+{{\sigma }_{z}}\left( \frac{1}{2}g\mu {{B}_{z}} \right)
\label{equ2}
\end{align}

\noindent
After diagonalizing S2, we obtain the spin splittings for different valley states, 

\begin{align}
{{E}_{{\textup{ZS}}(\pm )}}=2\left\lbrace {{(\frac{1}{2}g\mu {{B}_{z}})}^{2}}+{{\left( \frac{1}{2}g\mu {{B}_{x}}+{{\beta }_{\pm }}\left\langle {k_{x}^{\pm }} \right\rangle +{{\alpha }_{\pm }}\left\langle {k_{y}^{\pm }} \right\rangle  \right)}^{2}}+{{\left( \frac{1}{2}g\mu {{B}_{y}}-{{\beta }_{\pm }}\left\langle {k_{y}^{\pm }} \right\rangle -{{\alpha }_{\pm }}\left\langle {k_{x}^{\pm }} \right\rangle  \right)}^{2}}\right\rbrace^{\frac{1}{2}}
\label{equ3}
\end{align}

\noindent
Here, we replaced $\beta ,\alpha $ with ${{\beta }_{\pm }},{{\alpha }_{\pm }}$ to denote the two $v_{+}$ and $v_{-}$ valley states \cite{veldhorst_prbrap_2015,nestoklon_prb_2008}. Now, the expectation values of the momentum operator in a magnetic field $\mathbf{B}$ are,

\begin{align}
\left\langle {k_{x}^{\pm }} \right\rangle =\left\langle  {{\psi }_{v_{\pm} }} \right|-i\frac{\partial }{\partial x}-e\frac{{A_x}}{\hbar }\left| {{\psi }_{v_{\pm} }} \right\rangle \approx \left| e \right|\frac{\left\langle {A_{x}^{\pm }} \right\rangle }{\hbar }\\
\left\langle {k_{y}^{\pm }} \right\rangle =\left\langle  {{\psi }_{v_{\pm} }} \right|-i\frac{\partial }{\partial y}-e\frac{{A_y}}{\hbar }\left| {{\psi }_{v_{\pm} }} \right\rangle \approx \left| e \right|\frac{\left\langle {A_{y}^{\pm }} \right\rangle }{\hbar }
\label{equ4}
\end{align} 

\noindent
Since $\left\langle  {{\psi }_{{{v}_{\pm }}}} \right|-i\frac{\partial }{\partial x}\left| {{\psi }_{{{v}_{\pm }}}} \right\rangle \approx \left\langle  {{\psi }_{{{v}_{\pm }}}} \right|-i\frac{\partial }{\partial y}\left| {{\psi }_{{{v}_{\pm }}}} \right\rangle \approx 0$. For a magnetic field in the x-y plane, we assume, ${{A}_{z}}=0$, ${{A}_{x}}=z{{B}_{y}}$ and ${{A}_{y}}=-z{{B}_{x}}$, where $B_x$ and $B_y$ are the x and y components of the magnetic field respectively. Then,

\begin{align}
&\left\langle {k_{x}^{\pm}} \right\rangle \approx \left| e \right|\frac{\left\langle z \right\rangle }{\hbar }{{B}_{y}}\\
&\left\langle {k_{y}^{\pm}} \right\rangle \approx -\left| e \right|\frac{\left\langle z \right\rangle }{\hbar }{{B}_{x}}
\label{equ5}
\end{align} 

\noindent
Here, $\left\langle z \right\rangle$ is the dipole moment along [001] crystal direction. We assumed $\left\langle z_- \right\rangle=\left\langle z_+ \right\rangle=\left\langle z \right\rangle$. As the electrons in the $v_-$ and $v_+$ valley states have only $z$ valley component, we can replace $g$ here with ${{g}_{\bot }}$, the g-factor perpendicular to the valley axis\cite{roth_pr_1960}. So, for an in-plane magnetic field,

\begin{align}
{{E}_{{\textup{ZS}}(\pm )}}=2\left\lbrace {{\left( \frac{1}{2}{{g}_{\bot }}\mu {{B}_{x}}+{{\beta }_{\pm }}\left| e \right|\frac{\left\langle z \right\rangle }{\hbar }{{B}_{y}}-{{\alpha }_{\pm }}\left| e \right|\frac{\left\langle z \right\rangle }{\hbar }{{B}_{x}} \right)}^{2}}+{{\left( \frac{1}{2}{{g}_{\bot }}\mu {{B}_{y}}+{{\beta }_{\pm }}\left| e \right|\frac{\left\langle z \right\rangle }{\hbar }{{B}_{x}}-{{\alpha }_{\pm }}\left| e \right|\frac{\left\langle z \right\rangle }{\hbar }{{B}_{y}} \right)}^{2}}\right\rbrace^{\frac{1}{2}}
\label{equ6}
\end{align}

In the presence of only external magnetic field, $\mathbf{B}_{\textup{ext}}$, $B_x=B_{\textup{ext}}cos\phi$ and $B_y=B_{\textup{ext}}sin\phi$, with $\phi$ being the B-field angle relative to the [100] crystal orientation for a counter clockwise rotation of $B_{\textup{ext}}$. However, to be consistent with the angle $\theta$ defined in the experiment (i.e. an angle relative to the [110] crystal orientaiton with clockwise rotation of $\mathbf{B}_{\textup{ext}}$), we use the relation $\theta=45^{\circ}-\phi$. Now, using the definition of $B_x$ and $B_y$ from above, we can obtain the angular dependence of the spin splitting for different valley states,

\begin{align}
{{E}_{{\textup{ZS}}(\pm )}}={{g}_{\bot }}\mu B_{\textup{ext}}\left\lbrace{{1+4{{\left( \frac{\left| e \right|\left\langle z \right\rangle }{{{g}_{\bot }}\mu \hbar } \right)}^{2}}\left( {{\beta }_{\pm }}^{2}+{{\alpha }_{\pm }}^{2} \right)-4\frac{\left| e \right|\left\langle z \right\rangle }{{{g}_{\bot }}\mu \hbar }{{\alpha }_{\pm }}}}+{4\frac{\left| e \right|\left\langle z \right\rangle }{{{g}_{\bot }}\mu \hbar }{{\beta }_{\pm }}\sin 2\phi -8{{\left( \frac{\left| e \right|\left\langle z \right\rangle }{{{g}_{\bot }}\mu \hbar } \right)}^{2}}{{\beta }_{\pm }}{{\alpha }_{\pm }}\sin 2\phi}\right\rbrace^{\frac{1}{2}}
\label{equ7}
\end{align}

\noindent
Now, $1 \gg \left( \frac{\left| e \right|\left\langle z \right\rangle }{g\mu \hbar } \right)*{{\beta }_{\pm }}> \left( \frac{\left| e \right|\left\langle z \right\rangle }{g\mu \hbar } \right)*{{\alpha }_{\pm }}$ \cite{nestoklon_prb_2008}. So, we can ignore the second order terms. Thus simplifying equation (S9) we get,

\begin{align}
{{E}_{{\textup{ZS}}(\pm )}}={{g}_{\bot }}\mu {B_{\textup{ext}}}{{\left\{ 1-4\frac{\left| e \right|\left\langle z \right\rangle }{{{g}_{\bot }}\mu \hbar }{{\alpha }_{\pm }}+4\frac{\left| e \right|\left\langle z \right\rangle }{{{g}_{\bot }}\mu \hbar }{{\beta }_{\pm }}\sin 2\phi  \right\}}^{\frac{1}{2}}}
\label{equ8}
\end{align}

\noindent
After doing a series expansion and ignoring higher order terms, we can simplify this expression even further,

\begin{align}
{{E}_{{\textup{ZS}}(\pm )}}={{g}_{\bot }}\mu {B_{\textup{ext}}}\left\{ 1-2\frac{\left| e \right|\left\langle z \right\rangle }{{{g}_{\bot }}\mu \hbar }{{\alpha }_{\pm }}+2\frac{\left| e \right|\left\langle z \right\rangle }{{{g}_{\bot }}\mu \hbar }{{\beta }_{\pm }}\sin 2\phi  \right\}
\label{equ9}
\end{align}

\noindent
So, from this equation we can see that, {\bf{without the Dresselhaus contribution, there is no angular dependence or anisotropy in the spin splitting (or g-factor) for the different valley states}}. 
Now, for $\mathbf{B}_{\textup{ext}}$ along the $[110]$ and $[1\bar{1}0]$ crystal orientations, we get,

\begin{align}
& E_{_{{\textup{ZS}}(\pm )}}^{[110]}-{{g}_{\bot }}\mu {B_{\textup{ext}}}=2\left( {{\beta }_{\pm }}-{{\alpha }_{\pm }} \right)\left| e \right|\frac{\left\langle z \right\rangle }{\hbar }B_{\textup{ext}}\\
& E_{{\textup{ZS}}(\pm )}^{[1\overline{1}0]}-{{g}_{\bot }}\mu {B_{\textup{ext}}}=2\left( -{{\beta }_{\pm }}-{{\alpha }_{\pm }} \right)\left| e \right|\frac{\left\langle z \right\rangle }{\hbar }{B_{\textup{ext}}}
\label{equ10}
\end{align}

\noindent
Equation (S12) matches the analytic prediction of ref.\ \citen{veldhorst_prbrap_2015}. 
We can extract $\alpha_{\pm}$ and $\beta_{\pm}$ from the atomistic calculations as follows,

\begin{align}
& {{\alpha }_{\pm }}=-\frac{\left( E_{{\textup{ZS}}(\pm )}^{[110]}+E_{{\textup{ZS}}(\pm )}^{[1\bar{1}0]}-2{{g}_{\bot }}\mu {B_{\textup{ext}}} \right)}{4\left| e \right|\frac{\left\langle z \right\rangle }{\hbar }{B_{\textup{ext}}}} \\
&{{\beta }_{\pm }}=\frac{E_{{\textup{ZS}}(\pm )}^{[1\bar{1}0]}-E_{{\textup{ZS}}(\pm )}^{[1\bar{1}0]}}{4\left| e \right|\frac{\left\langle z \right\rangle }{\hbar }{B_{\textup{ext}}}}
\label{equ11}
\end{align}

\begin{figure}[htbp]
\includegraphics[]{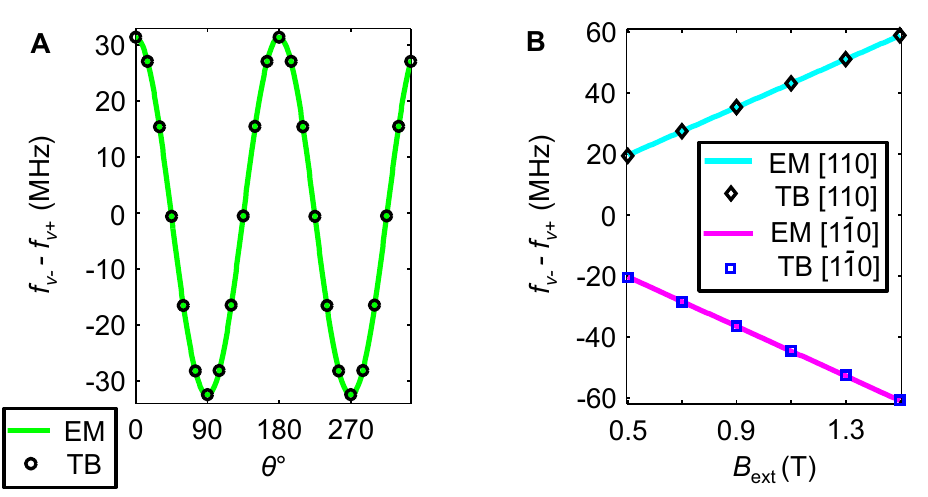}
\caption{{Effect of spin-orbit interaction: comparison between analytic effective mass and atomistic tight-binding calculations.} \textbf{A,} Comparison in angular dependence of $f_{v_-}-f_{v_+}$ for ${B_{\textup{ext}}}=$800mT. The angle of $\mathbf{B}_{\textup{ext}}$ is defined clockwise with respect to the $[110]$ crystal orientation) \textbf{B,} Magnetic field dependence of $f_{v_-}-f_{v_+}$ for ${\mathbf{B}_{\textup{ext}}}$ along $[110]$ and $[1\bar{1}0]$ crystal orientations. The tight-binding calculations correspond to the ideal (interface) case of Figs.\ 3e and 3f of the main text.}
\vspace{0cm}
\label{fi1}
\end{figure}

In Fig.\ S1, we compared $f_{v_-}-f_{v_+}$ calculated from this analytic model with the atomistic tight-binding results for the ideal interface case, shown in Figs.\ 3e and 3f of the main text. Both the anisotropic behavior of $f_{v_-}-f_{v_+}$ with respect to the angle of $\mathbf{B}_{\textup{ext}}$ (Fig.\ S1A), and the dependence on the magnitude ${B_{\textup{ext}}}$ along $[110]$ and $[1\bar{1}0]$ crystal orientations (Fig.\ S1B) show perfect agreement between analytic and atomistic calculations. From the atomistic calculations of Fig.\ 3a and equation (S14) and (S15), we extracted ${\beta }_{-}=10.117\times {{10}^{-15}}$eV$\cdot$m, ${{\beta }_{+}}=-9.3621\times {{10}^{-15}}$eV$\cdot$m, ${{\alpha }_{-}}=-1.2568\times {{10}^{-15}}$eV$\cdot$ m, ${{\alpha }_{+}}=-1.5920\times {{10}^{-15}}$eV$\cdot$ m for ${{g}_{\bot }}$=1.9937 and $\left\langle z \right\rangle=2.7925$ nm, calculated using $\left\langle z \right\rangle\approx1.5587 l_z$, where, ${{l}_{z}}={{\left( \frac{{{\hbar }^{2}}}{2{{m}_{l}}\left| e \right|{{E}_{z}}} \right)}^{\frac{1}{3}}}$ \cite{veldhorst_prbrap_2015}. Here, $E_{z}=6.77$ MVm$^{-1}$ is the vertical electric field.

To include the homogeneous fields from the micro-magnets, in equation (S3), we use,
${{B}_{x}}={B_{\textup{ext}}}\cos \theta +B_{micro}^{x}\left( \theta  \right)$, ${{B}_{y}}={B_{\textup{ext}}}\sin \theta +B_{\textup{micro}}^{y}\left( \theta  \right)$ and ${{B}_{z}}=B_{\textup{micro}}^{z}\left( \theta  \right)$. As, ${{B}_{x/y}}\gg{{B}_{z}}$ and $\left\langle x/y \right\rangle \gg \left\langle z \right\rangle$, we can still use equation (S6) and (S7) for $\left\langle {k_{x}^{\pm}} \right\rangle$ and $\left\langle {k_{y}^{\pm}} \right\rangle$.

\clearpage

\textbf{S2. Analytic model to explain the effect of gradient magnetic field on the valley dependent spin-splittings in a Si QD}

To develop an analytic model to capture the effect of the inhomogeneous magnetic field on the spin-splittings for different valley states, we assume the Hamiltonian in equation (S1) as our unperturbed Hamiltonian and include the magnetic field gradient as a perturbation,  

\begin{align}
H_{P}^{{\Delta \mathbf{B}}}=\frac{1}{2}g\mu {\bm{\sigma} .}{{{{\Delta \mathbf{B}}}}}=\frac{1}{2}g\mu \left\{ \sum\limits_{i}{{{\sigma }_{i}}\left( \sum\limits_{j}{\frac{dB_i}{dj}j} \right)} \right\}
\label{equ12}
\end{align}

\noindent
Here, $\frac{dB_i}{dj}$ are the magnetic field gradients along different directions, where $i,j$ correspond to x,y,z co-ordinates. Now the lowest 4 eigen-states of the unperturbed Hamiltonian are $\left| {{v}_{-}}\downarrow  \right\rangle $, $\left| {{v}_{-}}\uparrow  \right\rangle $, $\left| {{v}_{+}}\downarrow  \right\rangle $ and $\left| {{v}_{+}}\uparrow  \right\rangle $. The $1^{\textup{st}}$ order correction due to $H_{P}^{{\Delta \mathbf{B}}}$ to $\left| {{v}_{-}}\downarrow  \right\rangle $ is given by,

\begin{align}
\Delta {{E}_{{{v}_{-}}\downarrow }^{{{\Delta \mathbf{B}}}}}=\left\langle  {{v}_{-}}\downarrow  \right||\frac{1}{2}g\mu \left\{ \sum\limits_{i}{{{\sigma }_{i}}\left( \sum\limits_{j}{\frac{dB_i}{dj}j} \right)} \right\}\left| {{v}_{-}}\downarrow  \right\rangle 
\label{equ13}
\end{align}

To simplify our analytic calculation, we assume that the spin mixing due to SOI is very small and we can separate the spin part from the spatial part of the wavefunction. Using this approximation in \ref{equ13}, we get,

\begin{align}
\Delta {{E}_{{{v}_{-}}\downarrow }^{{{\Delta \mathbf{B}}}}}=\frac{1}{2}g\mu \left\{ \sum\limits_{i}{\left\langle  \downarrow  \right|{{\sigma }_{i}}\left| \downarrow  \right\rangle }\left( \sum\limits_{j}{\frac{d{{B}_{i}}}{dj}\left\langle j_- \right\rangle } \right) \right\}
\label{equ14}
\end{align}

\noindent
Here, $\left\langle j_- \right\rangle$ is the dipole moment along the $j$ direction, $\left\langle j_- \right\rangle=\left\langle  {{v}_{-}} \right|j\left| {{v}_{-}} \right\rangle $.
Now, for spins in an in-plane magnetic field, $\left\langle  \downarrow  \right|{{\sigma }_{x}}\left| \downarrow  \right\rangle =-\cos \phi $, $\left\langle  \downarrow  \right|{{\sigma }_{y}}\left| \downarrow  \right\rangle =-\sin \phi $ and $\left\langle  \downarrow  \right|{{\sigma }_{z}}\left| \downarrow  \right\rangle =0$.  The external magnetic field fully magnetizes the micro-magnets. So, the inhomogeneous magnetic field from the micro-magnets ($\frac{d{{B}_{i}}}{dj}$) depends on the direction ($\phi$) of the external magnetic field. Using these relations in equation (S19) we get,

\begin{align}
 \Delta {{E}_{{{v}_{-}}\downarrow }^{{{\Delta \mathbf{B}}}}}=-\frac{1}{2}g\mu \left\lbrace  \cos \phi \left( \sum\limits_{j}{\left\langle {{j}_{-}} \right\rangle \frac{dB_{x}^{\phi }}{dj}} \right)+\sin \phi \left( \sum\limits_{j}{\left\langle {{j}_{-}} \right\rangle \frac{dB_{y}^{\phi }}{dj}} \right) \right\rbrace
 \label{equ15}
\end{align}

\noindent
Now, for $\left| {{v}_{-}}\uparrow  \right\rangle $, $\left\langle  \uparrow  \right|{{\sigma }_{x}}\left| \uparrow  \right\rangle =\cos \phi $, $\left\langle  \uparrow  \right|{{\sigma }_{y}}\left| \uparrow  \right\rangle =\sin \phi $ and $\left\langle  \uparrow  \right|{{\sigma }_{z}}\left| \downarrow  \right\rangle =0$. So,

\begin{align}
\Delta {{E}_{{{v}_{-}}\uparrow }^{{{\Delta \mathbf{B}}}}}=\frac{1}{2}g\mu \left\lbrace \cos \phi \left( \sum\limits_{j}{\left\langle {{j}_{-}} \right\rangle \frac{dB_{x}^{\phi }}{dj}} \right)+\sin \phi \left( \sum\limits_{j}{\left\langle {{j}_{-}} \right\rangle \frac{dB_{y}^{\phi }}{dj}} \right) \right\rbrace
\label{equ16}  
\end{align}

\noindent
We can get similar expressions for $\left| v_+\downarrow \right\rangle$ and $\left|v_+\uparrow\right\rangle$. So, the change in spin splitting of both the valleys due to the gradient magnetic field is, 

\begin{align}
\Delta {{E}_{{\textup{ZS}}(\pm)}^{{{\Delta \mathbf{B}}}}}=g\mu  \left\lbrace \cos \phi \left( \sum\limits_{j}{\left\langle {{j}_{\pm}} \right\rangle \frac{dB_{x}^{\phi }}{dj}} \right)+\sin \phi \left( \sum\limits_{j}{\left\langle {{j}_{\pm}} \right\rangle \frac{dB_{y}^{\phi }}{dj}} \right) \right\rbrace 
\label{equ17} 
\end{align}

For an electron in an ideal (smooth) interface,  $\left\langle {{x}_{\pm}} \right\rangle \approx \left\langle {{y}_{\pm}} \right\rangle \approx 0$. But the presence of interface steps in realistic devices makes $\left\langle {{x}_{\pm}}\right\rangle$ and/or  $ \left\langle {{y}_{\pm}} \right\rangle$ non zero. We can ignore the 2nd order corrections because $\frac{1}{2}g\mu \frac{dB_i^{\phi}}{dj}\left\langle {{j}_{\pm}} \right\rangle \ll E_{\textup{{\textup{ZS}}}}$. Due to interface steps, $\left\langle {{x}_{-}}\right\rangle \neq \left\langle {{x}_{+}}\right\rangle$ and/or $\left\langle {{y}_{-}}\right\rangle \neq \left\langle {{y}_{+}}\right\rangle$, but $\left\langle {{z}_{-}}\right\rangle \approx \left\langle {{z}_{+}}\right\rangle$, for the electric field used in the experiment. So, in the presence of interface steps, the magnetic field gradient adds to the difference in ESR frequencies between the valley states, and from equation (S21), we get equation (3) of the main text.

\clearpage 

\begin{figure}[htbp]
\includegraphics[]{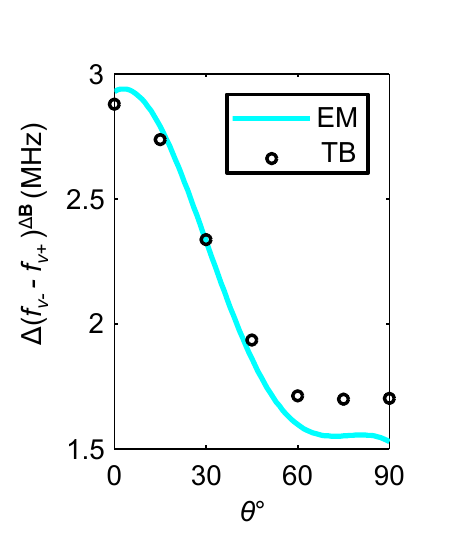}
\caption{Effect of the gradient magnetic field on the anisotropy of $f_{v_-}-f_{v_+}$: comparison between analytic effective mass and atomistic tight-binding calculations. For the latter, the contribution of the SOI is subtracted to obtain the change in $f_{v_-}-f_{v_+}$ due to ${\Delta \mathbf{B}}$ only from Fig. 1c of the main text (difference between green dashed line and black solid line).}
\vspace{0cm}
\label{fi1}
\end{figure}


In Fig.\ S2, we compared $\Delta \left({{f}_{{{v}_{-}}}}-{{f}_{{{v}_{+}}}}\right)^{{\Delta \mathbf{B}}}$ calculated using equation (2) of the main text with atomistic tight-binding calculations. We used $\left\langle {{x}_{-}}\right\rangle - \left\langle {{x}_{+}}\right\rangle\approx -0.17$ nm and $\left\langle {{y}_{-}}\right\rangle - \left\langle {{y}_{+}}\right\rangle = 0$ in equation (2). The atomistic results shown in Fig.\ S2 is the difference between the black curve (circular marker) and the green curve (square markers) shown in Fig.\ 1c. The analytic calculation qualitatively captures the tight-binding results. The little mismatch between the analytic and atomistic calculation is due to ignoring the spin mixing from SOI in our analytic model.
\clearpage  

\textbf{S3. Effect of the vertical electric field on the difference in ESR frequencies between valley states through both SOI and gradient magnetic field}

The vertical electric field, $E_z$ affects $f_{v_-}-f_{v_+}$, because it affects the SOI parameters\cite{nestoklon_prb_2008} and also the dipole moment parameters. The effect of $E_z$ on $f_{v_-}-f_{v_+}$ due to the SOI and gradient magnetic field, from atomistic tight-binding calculations, are shown in supplementary Figs.\ S3 and S4 respectively. Here $E_z$ only changes the magnitude of $f_{v_-}-f_{v_+}$, but not its sign. $E_z$ also effect the $B_{\textup{ext}}$ dependence of $f_{v_-}-f_{v_+}$ through SOI, as shown in supplementary Fig.\ S3B. Here also $E_z$ only changes the magnitude of the slope, $\frac{d\Delta(f_{v_-}-f_{v_+})^{\textup{SOI}}}{dB_{\textup{ext}}}$ but not its sign.

\begin{figure}[htbp]
\includegraphics[]{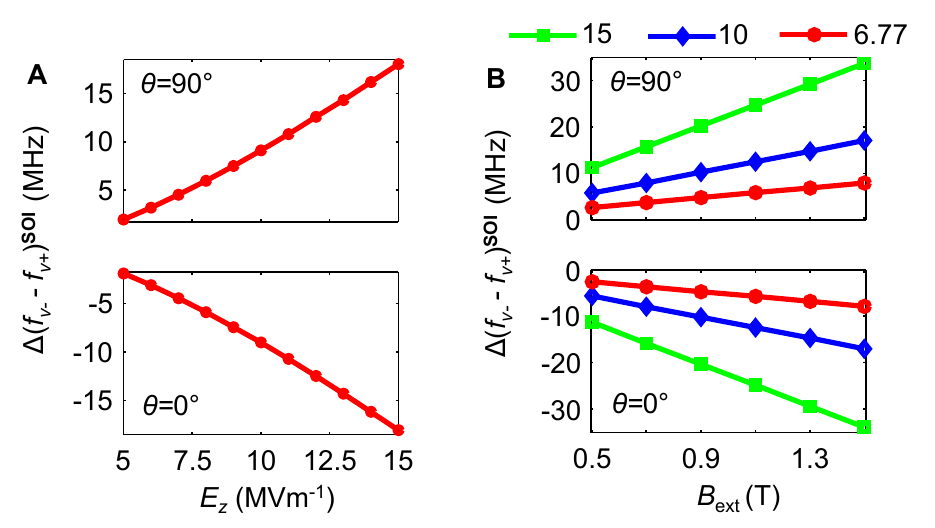}
\caption{Influence of the vertical electric field $E_z$ on the change in $f_{v_-}-f_{v_+}$ due to SOI. \textbf{A,} Increase in $\left|\Delta \left(f_{v_-}-f_{v_+}\right)^{\textup{SOI}}\right|$ with increasing $E_z$ for $B_{\textup{ext}}$=0.8 T. \textbf{B,} Change in $\Delta\left(f_{v_-}-f_{v_+}\right)^{\textup{SOI}}$ with changing $B_{\textup{ext}}$ for different $E_z$. $E_z$ changes the magnitude of the slope, $\left|\frac{d\Delta(f_{v_-}-f_{v_+})^{\textup{SOI}}}{dB_{\textup{ext}}}\right|$. Bottom panels in both Figs.\ A and B corresponds to $\mathbf{B}_{\textup{ext}}$ along $[110]$ ($\theta=0^{\circ}$) and in top panels $\mathbf{B}_{\textup{ext}}$ is along $[1\bar{1}0]$ ($\theta=90^{\circ}$). Interface condition and parabolic confinement for the dot used in these simulations are the same as that used to match experimental data in Figs.\ 1 and 2 of main text.}
\vspace{0cm}
\label{fi3}
\end{figure}

\clearpage 
\begin{figure}[htbp]
\includegraphics[]{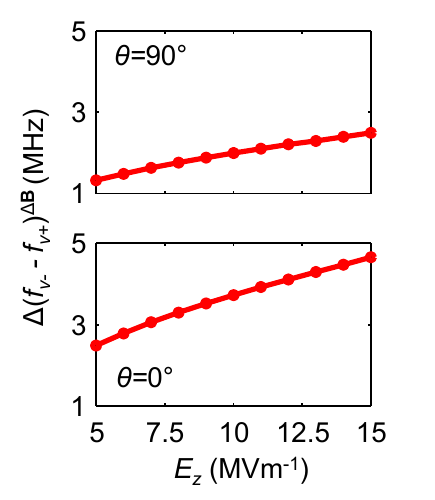}
\caption{Influence of the vertical electric field $E_z$ on the change in $f_{v_-}-f_{v_+}$ due to the gradient magnetic field ${\Delta \mathbf{B}}$, for $\mathbf{B}_{\textup{ext}}$ along $[110]$ $\theta=0^{\circ}$ (bottom panel) and $[1\bar{1}0]$ ($\theta=90^{\circ}$) (top panel). Interface condition and parabolic confinement for the dot used in these simulations are the same as that used to match experimental data in Figs.\ 1 and 2 of the main text. Increasing $E_z$ increases $\left|\Delta\left(f_{v_-}-f_{v_+}\right)^{{\Delta \mathbf{B}}}\right|$.}
\vspace{0cm}
\label{fi4}
\end{figure}



\clearpage 

\textbf{S4. Modeling the stray magnetic field induced by the micro-magnets}

We calculated the local magnetic field created by the micro-magnets when the external magnetic field is applied along the y$'$ ([110]) ($\theta=0^\circ$) or along the x$'$ ([1$\bar{1}$0]) axis ($\theta=90^\circ$) of the device picture in Fig. S5, assuming that the micro-magnets are fully magnetized \citep{Goldman2000,Pioro-Ladriere2007}. The shape of the micro-magnets is shown in ref.\ \citen{Kawakami2014}. 

\begin{figure}[htbp]
\includegraphics[]{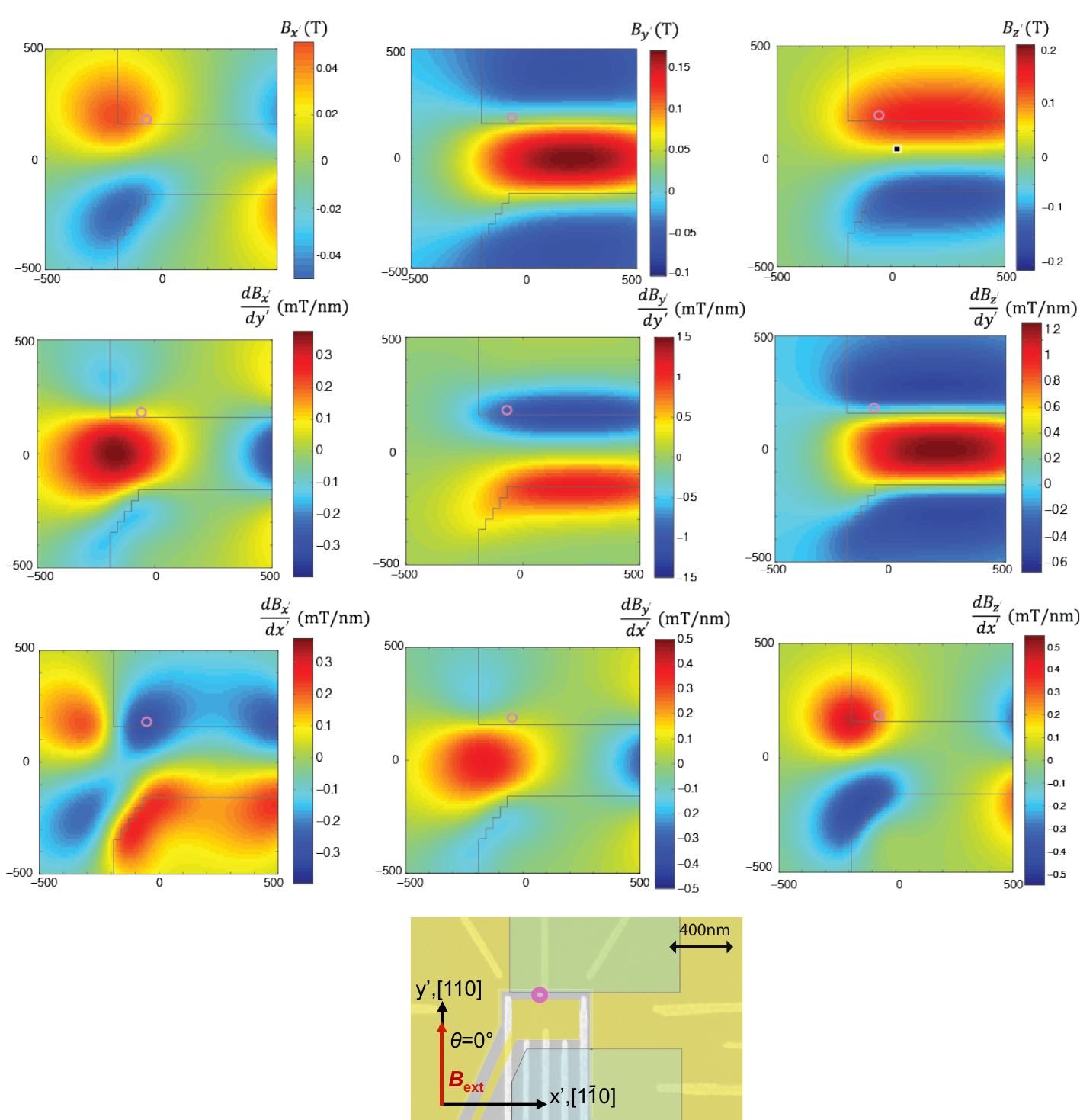}
\caption{Numerically computed x$'$, y$'$ and z components of the magnetic field and their gradients along the x$'$ and y$'$ axes induced by the micro-magnets when the external magnetic field is applied along the y$'$ axis in the plane of the Si quantum well, for fully magnetized micro-magnets. The black solid lines indicate the edges of the micro-magnet as simulated.}
\vspace{0cm}
\label{fig:Simu:BextZ}
\end{figure}

\begin{figure}[htbp]
\includegraphics[]{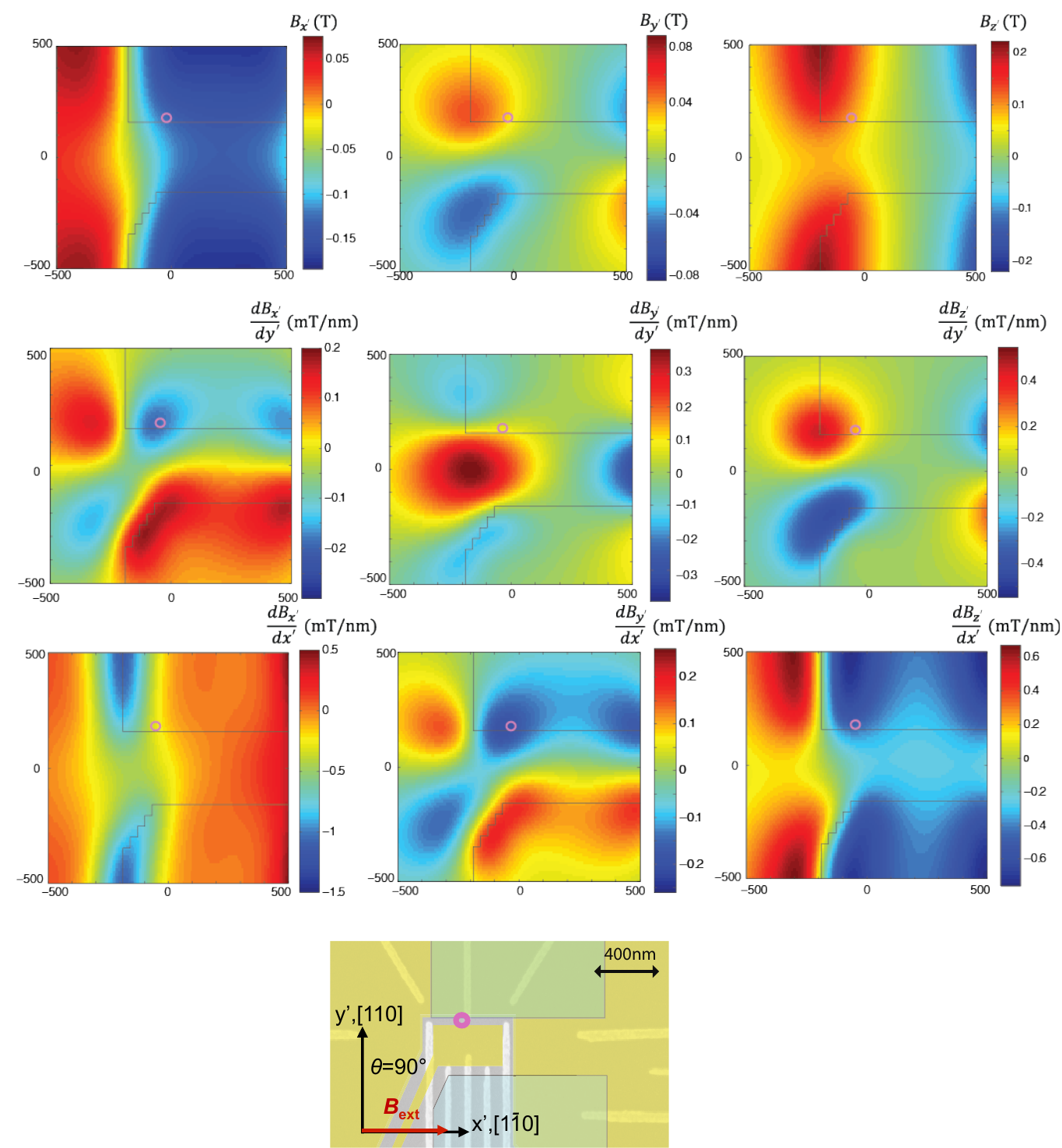}
\caption{The same as in Fig.~\ref{fig:Simu:BextZ} when the external magnetic field is applied along the x$'$ axis.}
\vspace{0cm}
\label{fig:Simu:BextX}
\end{figure}

Fig.~\ref{fig:Simu:BextZ} and Fig.~\ref{fig:Simu:BextX} show the results of the numerical calculation of the total magnetic field gradient when the external magnetic field is applied along the y$'$ axis and along the x$'$ axis respectively, and the pink circle shows the estimated dot position.

We calculated the stray magnetic field created by the micro-magnets when the external magnetic field is applied between the y$'$ axis and the x$'$  axis ($0^\circ < \theta < 90^\circ$) according to the following approximation,

\begin{equation}
B_i^{\theta}=\cos^2\theta B_i^{0^\circ}+\sin^2 \theta B_i^{90^\circ},
\end{equation}

\noindent
where $i=$x$'$,y$'$,z. 

We ignore  $\frac{dB_i^{\theta}}{dz}$, with $i=$x$'$,y$'$,z.  For small electric field like $E_z$=6.77 MVm$^{-1}$ used to match the experiment, the electron wavefunctions of the two valley states have similar spread along z, so $\left\langle  {{v}_{-}} \right|\frac{dB_{i}^{\theta }}{dz}z\left| {{v}_{-}} \right\rangle \approx \left\langle  {{v}_{+}} \right|\frac{dB_{i}^{\theta }}{dz}z\left| {{v}_{+}} \right\rangle $. Thus the effects of $\frac{dB_i^{\theta}}{dz}$ on $f_{v-}-f_{v+}$ should be small. However, the effects of these $\frac{dB_i^{\theta}}{dz}$ on $f_{v-}$ or $f_{v+}$ can be larger, but as shown in the Figs.\ 1d and 2b of the main text, the effect of the gradient magnetic field on $f_{v_{\pm}}$ is negligible compared to that of the homogeneous magnetic fields. 

The micro-magnetic field used in our calculations are $\mathbf{B}_{\textup{micro}}^{0^{\circ}}=\left(0.03,0.03,0.139\right)$ T, $\mathbf{B}_{\textbf{micro}}^{90^{\circ}}=\left(-0.122,0.03,0.089\right)$ T, $\frac{d\bm{B}^{0^{\circ}}}{dx'}=\left(-0.185,0.056,0.217\right)$ mT/nm, $\frac{d\bm{B}^{0^{\circ}}}{dy'}=\left(-0.0653,-0.93,-0.052\right)$ mT/nm, $\frac{d\bm{B}^{90^{\circ}}}{dx'}=\left(-0.272,-0.185,-0.456\right)$ mT/nm, $\frac{d\bm{B}^{0^{\circ}}}{dy'}=\left(-0.184,-0.065,0.211\right)$ mT/nm.

\clearpage 

\clearpage
\begin{figure}[htbp]
\includegraphics[]{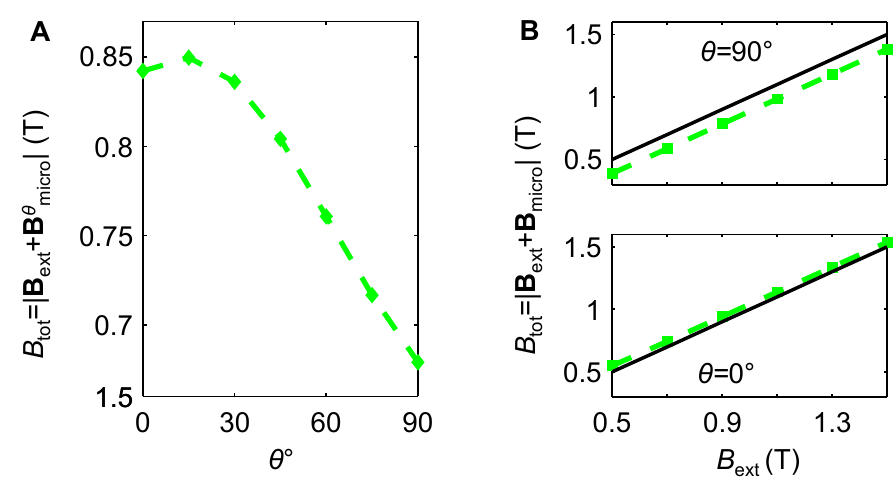}
\caption{Effect of $\mathbf{B}_{\textup{micro}}$ on the total homogeneous magnetic field. \textbf{A,} Anisotropic total magnetic field, due to the presence of the homogeneous magnetic field from the micro-magnet, $\mathbf{B}_{\textup{micro}}^{\theta}$.  $\mathbf{B}_{\textup{micro}}^{\theta}$ adds vectorially to $\mathbf{B}_{\textup{ext}}$. Also the magnetization of the micro-magnet depends on the direction of $\mathbf{B}_{\textup{ext}}$ or $\theta$. So $\mathbf{B}_{\textup{micro}}^{\theta}$ itself is anisotropic, hence the superscript $\theta$. \textbf{B,} Total magnetic field (shown in green dashed line with square markers) along   along $[110]$ ($\theta=0^{\circ}$) (bottom panel) and $[1\bar{1}0]$ ($\theta=90^{\circ}$) (top panel) with changing $B_{\textup{ext}}$. The black line represents $B_{\textup{tot}}=B_{\textup{ext}}$. }
\vspace{0cm}
\label{fi4}
\end{figure}

\clearpage
\textbf{S5. Extracting SOI parameters from the experimental measurements}

The contributions of the SOI due to $B_{\textup{ext}}$ on $f_{v_-}-f_{v_+}$ is given by equation 1 of the main text. The dependence of $f_{v_-}-f_{v_+}$ on $B_{\textup{ext}}$ only comes from this equation. Thus we can calculate $\frac{d(f_{v_-}-f_{v_+})}{dB_{\textup{ext}}}$,

\begin{align}
\frac{d(f_{v_-}-f_{v_+})}{dB_{\textup{ext}}^{\phi}} \approx
\frac{4\pi \left| e \right|\left\langle z \right\rangle }{{{h}^{2}}} \left\lbrace\left( {{\beta }_{-}}-{{\beta }_{+}} \right)\sin 2\phi -\left( {{\alpha }_{-}}-{{\alpha }_{-}} \right) \right\rbrace
\label{equ18}
\end{align}
Now, for $\mathbf{B}_{\textup{ext}}$ along the $[110]$ and $[1\bar{1}0]$ crystal orientations, we get,
\begin{align}
& \frac{d(f_{v_-}-f_{v_+})}{dB_{\textup{ext}}^{[110]}} \approx
\frac{4\pi \left| e \right|\left\langle z \right\rangle }{{{h}^{2}}} \left\lbrace\left( {{\beta }_{-}}-{{\beta }_{+}} \right) -\left( {{\alpha }_{-}}-{{\alpha }_{-}} \right) \right\rbrace \\
& \frac{d(f_{v_-}-f_{v_+})}{dB_{\textup{ext}}^{[1\bar{1}0]}} \approx
\frac{4\pi \left| e \right|\left\langle z \right\rangle }{{{h}^{2}}} \left\lbrace-\left( {{\beta }_{-}}-{{\beta }_{+}} \right) -\left( {{\alpha }_{-}}-{{\alpha }_{-}} \right) \right\rbrace
\label{equ19}
\end{align}

Then we can extract $\left( {{\beta }_{-}}-{{\beta }_{+}} \right)$ and $\left( {{\alpha }_{-}}-{{\alpha }_{-}} \right)$,

\begin{align}
& \left( {{\beta }_{-}}-{{\beta }_{+}} \right)\approx \frac{{{h}^{2}}}{8\pi \left| e \right|\left\langle z \right\rangle }\left\{ \frac{d({{f}_{{{v}_{-}}}}-{{f}_{{{v}_{+}}}})}{dB_{\textup{ext}}^{[110]}}-\frac{d({{f}_{{{v}_{-}}}}-{{f}_{{{v}_{+}}}})}{dB_{\textup{ext}}^{[1\bar{1}0]}} \right\} \\
& \left( {{\alpha }_{-}}-{{\alpha }_{+}} \right)\approx -\frac{{{h}^{2}}}{8\pi \left| e \right|\left\langle z \right\rangle }\left\{ \frac{d({{f}_{{{v}_{-}}}}-{{f}_{{{v}_{+}}}})}{dB_{\textup{ext}}^{[110]}}+\frac{d({{f}_{{{v}_{-}}}}-{{f}_{{{v}_{+}}}})}{dB_{\textup{ext}}^{[1\bar{1}0]}} \right\}
\label{equ20}
\end{align}
\vspace{-0.5cm}

Thus from the experimentally measured $\frac{d(f_{v_-}-f_{v_+})}{dB_{\textup{ext}}}$ along the $[110]$ and $[1\bar{1}0]$ crystal orientations, we extract the SOI parameters.
\clearpage

\textbf{S6. Steps to calculate Rabi frequency for electric-dipole spin resonance (EDSR) due to SOI and inhomogeneous B-field}

The time dependent perturbation to the Hamiltonian due to an ac electric field $E_{\textup{ac}}(t)=E_{\textup{ac}}f(t)$ is $H_{\textup{p}}(t)=qE_{\textup{ac}}(t)x$. Then the off-diagonal matrix element between up and down spin states for both the valley states is $ \left\langle{{v}_{\pm}}\downarrow\right| H_{\textup{p}}(t)  \left|{{v}_{\pm}}\uparrow \right\rangle = qE_{\textup{ac}}f(t)\left\langle{{v}_{\pm}}\downarrow\right| x \left|{{v}_{\pm}}\uparrow \right\rangle$. For a sinusoidal perturbation, $f(t)=\cos(wt)$, where $w=2\pi f_{v_{\pm}}$, the Rabi frequency is given by, 
\begin{align}
f_{Rabi}(v_{\pm})=\frac{qE_{\textup{ac}}|\left\langle{{v}_{\pm}}\downarrow\right| x \left|{{v}_{\pm}}\uparrow \right\rangle|}{h}
\label{equ21}
\end{align}
Using atomistic tight-binding wavefunctions we calculate $|\left\langle{{v}_{\pm}}\downarrow\right| x \left|{{v}_{\pm}}\uparrow \right\rangle|$ and then use \ref{equ21} to calculate the Rabi frequency. Here we use\cite{wister_prb_2017}, $E_{\textup{ac}}=2$ kVm$^{-1}$. Since interface roughness affects $|\left\langle{{v}_{\pm}}\downarrow\right| x \left|{{v}_{\pm}}\uparrow \right\rangle|$ in a Si QD, these Rabi frequencies will strongly depend on the interface condition.